\documentclass[aps,prd,reprint,showpacs,nofootinbib,twoside,notitlepage]{revtex4-1} 
\usepackage{graphicx}
\usepackage{euscript,amssymb}
\usepackage{amsfonts}
\usepackage{amssymb}
\usepackage{amsbsy}
\usepackage{amsmath}
\usepackage{nccmath}   
\usepackage{color}
\usepackage{bbold}

\usepackage[caption=false]{subfig} 

\newcommand{\be}{\begin{equation}}
  \newcommand{\ee}{\end{equation}}
\newcommand{\ben}{\begin{eqnarray*}}
  \newcommand{\een}{\end{eqnarray*}}
\newcommand{\bea}{\begin{eqnarray}}
  \newcommand{\eea}{\end{eqnarray}}
\newcommand{\bdm}{\begin{displaymath}}
  \newcommand{\edm}{\end{displaymath}}
\newcommand{\ba}{\begin{align}}
  \newcommand{\ea}{\end{align}}
\newcommand{\lb}{\label}

\newcommand{\sgn}{\text{sgn}\,}

\begin{document}

\title{Singularity avoidance for collapsing quantum dust in the
  Lema\^{\i}tre-Tolman-Bondi model} 

\author{Claus Kiefer}

\email{kiefer@thp.uni-koeln.de}

\author{Tim Schmitz}

\email{tschmitz@thp.uni-koeln.de}

\affiliation{Institut f\"ur Theoretische Physik, Universit\"{a}t zu
K\"{o}ln, Z\"{u}lpicher Stra\ss e 77, 50937 K\"{o}ln, Germany}

\date{\today}

\begin{abstract}

  We investigate the fate of the classical singularity in a collapsing
  dust cloud. For this purpose, 
we quantize the marginally bound Lema\^{i}tre-Tolman-Bondi model for
spherically-symmetric dust collapse by considering each dust shell 
in the cloud individually, taking the outermost shell as a
representative.
Because the dust naturally provides a preferred notion of
time, we can construct a quantum mechanical model for this shell
and demand unitary evolution for wave packets. It turns out that the
classical singularity can generically be avoided provided the
quantization ambiguities fulfill some weak conditions. We
demonstrate that the collapse to a singularity is replaced by a
bounce followed by an expansion. We finally construct a quantum corrected
spacetime describing 
bouncing dust collapse and calculate the time from collapse to expansion. 
	
\end{abstract}

\maketitle

\section{Introduction}

It is an open problem whether the ubiquitous singularities of general
relativity will disappear after quantization. Since there is no
consensus so far on the appropriate quantum theory of gravity, this
question can be decided only within a given approach and for certain
classes of models. 

In this paper, we shall address the fate of the classical singularity
for a collapsing dust cloud. The framework will be quantum
geometrodynamics, which is the canonical formulation based on metric
variables. Although this 
approach may not be the most fundamental one, it is a conservative
approach: one can arrive at the quantum constraint equations by
devising wave equations from which the classical Einstein equations
follow in the semiclassical (WKB) limit \cite{OUP}.

There already exist various results on the fate of singularities for
collapsing spherically-symmetric dust {\em shells}. Using an effective
one-loop action with 
an Einstein-Hilbert term plus a Weyl tensor-squared term, it was found
that a thin null dust shell collapses and re-expands instead of ending
in a black-hole (BH) singularity \cite{FrolovNullShell}.
In quantum geometrodynamics, the quantization of a collapsing dust
shell was discussed in a mathematically rigorous way in 
\cite{HajicekQuantumNullShells,HajicekKieferNullShells,AmbrusHajicekLifetime},
see also \cite{HajicekNullShells} for a review.  The demand for a
unitary evolution leads to a wave {\em vanishing} at the
origin, that is, at the place (more precisely, the time) where
classically the singularity sits.  The shell, if represented by a wave
packet, collapses to a minimal radius inside its horizon and then
re-expands. In the classical theory, this re-expanding wave packet corresponds
to a white hole. That the singularity is avoided in this way is
not surprising. In a unitary time evolution it is not possible that
the wave packet {\em disappears} in a singularity -- it must re-expand.

A different but related situation arises for quantum cosmological
models. There, unitarity does not hold
for the standard Wheeler-DeWitt equation \cite{OUP}. It is, however,
possible to impose the `DeWitt criterion' of vanishing wave function
in the limit of approaching the classical cosmological singularity. This was
investigated for several models; see, for example, \cite{Bouhmadi16}
and the references therein. Recently, the DeWitt criterion was
generalized in order to accommodate the conformal nature of the
configuration space \cite{KKP19}.

Concerning the fate of collapsing dust shells, there are also
investigations in other approaches, notably from loop quantum gravity
\cite{RovelliPlanckStars,AshtekarCosmology,AshtekarEvaporation,BambiBounce}.
Again, collapsing quantum shells turn into expanding ones.
A major issue there is the question of the lifetime of the BH-like temporary object and the behavior of the horizon. This is of great importance for relating theses scenarios to potential
observations. They provide realistic models only if the lifetime is bigger than
the current age of our Universe. Otherwise, they cannot be applied to
describing the quantum collapse of astrophysical objects such as
supernovae. 
 
Concerning the details of the scenario, there are a variety of ideas
available: the 
horizon could, for example, disappear during the bounce
\cite{BambiBounce,MalafarinaBounce,BarceloBounce3,BarceloBounce2} or could
be in a superposition of BH and white hole (WH) horizons, with a smooth
transition between the two in the form of a `grey
horizon' \cite{HajicekQuantumNullShells}.
There have also been different pictures about the detailed mechanism
that leads to the quantum effects at the horizon, a spacetime region
in which the curvature is usually low.
 Haggard and Rovelli, for example, 
envision an accumulation of quantum effects over time
\cite{HaggardRovelliBounce}, while Barcel\'{o} \textit{et al.} propose
a shockwave propagating outward from the would-be singularity
\cite{BarceloBounce3,BarceloBounce2}. There is also little
consensus about the lifetime, different approaches to the problem
giving different results  
\cite{AmbrusHajicekLifetime,BarceloLifetime,ChristodoulouLifetime2,ChristodoulouLifetime}.
A recent review is given in \cite{MalafarinaBounceRev}.

In this paper, we shall discuss these problems for the inhomogeneous
spherically-symmetric dust collapse described by the
Lema\^{i}tre-Tolman-Bondi (LTB) model; see, for example,
\cite{Krasinski97} for a presentation of the classical LTB model.
Its quantization in the geometrodynamical context was presented
in \cite{VazMargLTB1,KieferLTB1}. While it was possible in this model to recover
Hawking radiation \cite{VazMargLTB3}, to compute non-thermal
corrections to it \cite{KieferLTB2,KieferLTB3}, and to
investigate BH entropy and the BH mass spectrum \cite{VazMargLTB2},
the question of singularity avoidance could not be settled.
The main reason for this failure is the inhomogeneous nature of a dust
cloud and the ensuing functional form of the quantum constraints.
Similarly,
while it was claimed that in spherically-symmetric loop quantum
gravity the singularity is avoided due to the fundamentally
quantized nature of space
\cite{BojowaldLoopSingularities,HusainLoopSingularities},
investigating different loop quantum gravity inspired corrections to
the LTB model has not suggested any particular mechanism for this
avoidance; a singularity seems to form just as it does classically
\cite{BojowaldLoopLTB,BojowaldLoopLTB2}. 

Here, we shall develop a different approach to quantizing the
LTB model. The idea is to consider each shell individually, sidestepping
some technical and conceptual difficulties, and try to infer the
behavior of the full dust cloud from our results. This will enable us
to tackle the question of singularity avoidance and to suggest a
scenario with a bounce
as the typical behavior of the quantized dust cloud. Singularities can
thus be avoided. This bounce is a direct consequence of the unitary
evolution with respect to dust proper time.

Our paper is organized as follows. 
In Sec.\ \ref{ch:chapter_2} we introduce the reader to
the LTB model and lay the classical foundations for our approach. We 
then develop and investigate the corresponding quantum theory in Sec.\
\ref{ch:chapter_3}, first making general statements about its states,
and then examining a specific one in the form of a wave packet. Based
on the dynamics of this wave packet, we construct a quantum
corrected space time for dust collapse and discuss some of its
properties in Sec.\ \ref{ch:chapter_4}. We discuss, in particular, the
lifetime for the wave packet to collapse and re-expand. Sec.~V contains
our conclusions.  


\section{The classical LTB model and its on-shell action}
\label{ch:chapter_2}
We give here a brief introduction to the LTB model. It
is a spherically-symmetric solution of the Einstein
equations with non-rotating dust of mass density $\epsilon$ as its
source. Its line element reads
\begin{align}
  &ds^2=-c^2d\tau^2+\frac{ R'^2}{1+2f}~d\rho^2+R^2~d\Omega^2\,,
    \label{eq:ltb_metric}\\
  &\text{with}\quad \frac{8\pi G}{c^2}\epsilon=\frac{F'}{R^2R'}
    \quad\text{and}\quad
        \frac{\dot{R}^2}{c^2}= \frac{F}{R}+2f. \label{eq:eom_ltb} 
\end{align}
A prime (dot) denotes a derivative with respect to $\rho$
($\tau$). The cosmological constant is set to zero. For the
time coordinate one chooses the dust proper time $\tau$ and for the
radial coordinate the variable $\rho$, which continuously labels the spherically
symmetric dust shells at fixed $\tau$. In the following, we shall set
$G=1=c$.  

In these units, $F(\rho)$ is twice the Misner-Sharp mass
(see e.g. \cite{Szabados09}, p.~40) for the LTB
spacetime, which gives the active gravitating mass that is contained
in the shell 
with label $\rho$. From the condition $R(\tau,\rho)=F(\rho)$ one
can also infer whether a shell coincides with an
apparent horizon; the
horizon can be future or past depending on the sign of $\dot{R}$. 
The energy function $f(\rho)$ plays a role for
the general LTB model, but for simplicity we will in the following
restrict ourselves to the marginally bound LTB model for which $f=0$. 

An important quantity is $R(\tau,\rho)$, which is the 
curvature radius of the shell labelled by $\rho$ at time $\tau$; it describes
how the dust shells collapse or expand. A central or shell
focusing singularity forms in the LTB model when shells collapse to the
point $R=0$. 

In addition to the central singularity also shell crossing
singularities can appear. They occur when two dust shells occupy
the same radius, that is, when $R'=0$. They are generally assumed to be an artifact
of using a simplistic matter model and hence are considered unphysical. 
We will not address these singularities here,
because one can choose initial conditions such that they do not occur.
Moreover, it is possible to extend the spacetime beyond them; see
\cite{NolanShellCrossing,SeifertShellCrossing} and the references
therein. 

We emphasize that the equation of motion relevant for $R$, the second equation
in \eqref{eq:eom_ltb}, only depends on $R$ and $F$ (and also on $f$
for the non-marginally bound case), but not on their spatial
derivatives. When a mass function is given, different dust shells
are decoupled, as they do not dynamically influence each other.

Based on this decoupling, we can consider the different shells in the
LTB model independently. Consequently, we will quantize a single shell
in the marginally bound LTB model and then try to deduce the dynamics
of the full dust cloud. In the following, we will derive a Hamiltonian
for the outermost dust shell.  

We start from the Einstein-Hilbert action
\begin{multline}
  \label{EH-action}
S=\frac{1}{16\pi}\int_\mathcal{M}
d^4x~\sqrt{-g}\,{\mathcal R}[g]\\+\frac{1}{8\pi}\int_{\partial\mathcal{M}}
d^3x~\eta\sqrt{|h|}\,\left(k-k^0\right), 
\end{multline}
and insert a marginally bound LTB solution in the coordinates
$(\tau,\rho,\theta,\phi)$, where the angular coordinates can be
integrated out immediately. In \eqref{EH-action}, $k$ is the trace of the
extrinsic curvature of $\partial\mathcal{M}$, and $k^0$ is the same
quantity for the case that this hypersurface is embedded into flat space. The
factor $\eta$ is equal to $1$ when $\partial\mathcal{M}$ is timelike
and $-1$ when it is spacelike \cite{PoissonRelativistsToolkit}. 

For the boundary $\partial\mathcal{M}$ of the spacetime
$\mathcal{M}$ we take 
$\partial\mathcal{M}=\mathcal{B}_o\cup\mathcal{B}_{\tau_1}\cup\mathcal{B}_{\tau_2}$,
where $\mathcal{B}_{\tau_{1/2}}$ are spacelike hypersurfaces of fixed
constant dust proper time with $\tau_1<\tau_2$, and $\mathcal{B}_o$ is
the timelike boundary coinciding with the worldtube of the outermost
dust shell $\rho=\rho_o$. We will mostly not concern ourselves with
the geometry outside the cloud, although one can always attach a
Schwarzschild exterior with mass
$M=\frac{1}{2}F(\rho_o)=:\frac{1}{2}F_o$. Below we will refer to 
$F_o$ (twice the mass contained in the outermost shell) as twice the
ADM energy of the dust cloud, $2E_{\text{ADM}}$, always with this
exterior geometry in mind.  

Taking the trace of the Einstein equations for LTB gives
\begin{align*}
\sqrt{-g}\,{\mathcal R}[g]=8\pi\epsilon\, R^2R'\sin\theta=F'\sin\theta,
\end{align*}
where we have used the equations of motion
\eqref{eq:eom_ltb}. This gives for the bulk part of the action \eqref{EH-action},
$S_\mathcal{M}$, the expression
\begin{equation*}
S_\mathcal{M}=\frac{1}{4}\int
d\tau\int_{0}^{\rho_o}d\rho~F'=\frac{1}{4}\int
d\tau~F_o=\frac{1}{4}\int d\tau~R_o\dot{R}_o^2,
\end{equation*}
where $R_o$ denotes the radius of the outermost shell. 
We have made here the assumption that the innermost shell
contains no mass, $F(0)=0$, and have used the remaining part of
\eqref{eq:eom_ltb}. 
 
Now we turn to the boundary terms. Calculating the trace of the extrinsic curvature of the timelike boundary $\mathcal{B}_o$ gives
\begin{align*}
k_o=\frac{2}{R}.
\end{align*} 
Since this matches the trace of the extrinsic curvature for the same
hypersurface embedded into flat space, $k_o^0$, the
corresponding boundary term in \eqref{EH-action} vanishes. Note that the same
would hold for a boundary term at the innermost shell $\rho=0$. 

Let us now calculate the contributions from the temporal
boundaries. The trace of the extrinsic curvature of the
$\tau=\text{const.}$ hypersurfaces is given by 
\begin{equation*}
k_\tau=\frac{\dot{R}'}{R'}+2\frac{\dot{R}}{R},
\end{equation*}
while $k^0_\tau$ simply vanishes. This gives
\begin{equation*}
S_\mathcal{B_{\tau}}=-\frac{1}{2}\int_{0}^{\rho_o}d\rho~\left( R^2\dot{R}'+2RR'\dot{R}\right) =-\frac{1}{2}\left[ R^2\dot{R}\right]^{\rho_o}_{0}.
\end{equation*}
Combining the two terms for $\tau_1$ and $\tau_2$ gives a more
convenient form for these boundary contributions. One has to keep in
mind that the normal to 
$\mathcal{B}_{\tau_1}$ is future-directed, while the normal to
$\partial\!\mathcal{M}$ is past-directed in the region
$\mathcal{B}_{\tau_1}$. The past-directed boundary term hence carries
an additional sign $-1$ \cite{PoissonRelativistsToolkit}, giving 
\begin{align*}
\left. S_{\mathcal{B}_{\tau}}\right|^{\tau_2}_{\tau_1} &=-\frac{1}{2}\left[ \left. R^2\dot{R}\right|_{\tau_2}-\left. R^2\dot{R}\right|_{\tau_1}\right] ^{\rho_o}_{0}\\
&=-\frac{1}{2}\int d\tau~ \frac{\partial}{\partial\tau}\left[ R^2\dot{R}\right] ^{\rho_o}_{0}\\
&=-\frac{3}{4}\int d\tau~ \left[R\dot{R}^2\right]
                                                                                                           ^{\rho_o}_{0}=-\frac{3}{4}\int d\tau ~R_o\dot{R}_o^2. 
\end{align*}
Here we have used $R^2\ddot{R}=-\frac{1}{2}R\dot{R}^2$, which follows from the time independence of $F=R\dot{R}^2$.

The full action for an LTB solution of the outermost shell then reads
\begin{equation}
  \lb{S-full}
S=-\frac{1}{2}\int d\tau ~R_o\dot{R}_o^2. 
\end{equation}
We note that choosing Brown-Kucha\v{r} dust as the matter component,
 the dust action trivially vanishes on-shell
\cite{KucharBrownDust}. 

 We have now arrived at an action that describes the dynamics of the
 outermost shell. This is not surprising, since we have already
 inserted the proper dynamics for the dynamical field $R(\rho)$ and
 are now left with a prescription for how the boundary conditions
 given at the initial time $\tau_1$ are to be evolved into the
 future. In this sense, the above action \eqref{S-full} is an action
 for the outermost shell on the background of all other shells. 

We note that including the boundary terms has only
contributed to the prefactor of the action. If we neglected them,
we would only find a different prefactor that would leave the
classical dynamics unchanged and would only introduce minor changes to
the quantum model below.

The momentum conjugate to $R_o$
and the Hamiltonian corresponding to \eqref{S-full} then read, respectively,
\begin{align}
P_o&=-R_o\dot{R}_o,\\
H&=-\frac{P_o^2}{2R_o}.\label{eq:inf_h}
\end{align}
This Hamiltonian is the negative of the ADM energy,
\bdm
H=-\frac{1}{2}R_o\dot{R}^2_o=-\frac{1}{2}F_o=-E_{\text{ADM}},
\edm
implying its conservation. It is then obvious that
$H$ gives the expected dynamics. Adjusting the constant
of motion $F_o$, this Hamiltonian describes the dynamics of any single
shell in the LTB model, not just the outermost one. It is also consistent with
the on-shell Hamiltonian constraint for a marginally bound LTB model,
see \cite{BojowaldLoopLTB}.  

The fact that the Hamiltonian \eqref{eq:inf_h} is negative, although
surprising at first glance, reflects the fact that the gravitational
kinetic term in the Hamiltonian constraint is not positive definite (a
feature that can be related to the attractivity of gravity
\cite{GK94}). As we have seen above, it is possible here to recover a
positive notion of energy from it. A
similar observation was made in 
\cite{GieselLTB}, where phantom dust had to be used to recover a
positive Hamiltonian for the LTB model.

We note that it is not possible to arrive at an action for
non-marginally bound LTB models in a similar way, but an effective Hamiltonian is
easily constructed by simply adding a potential term $fR$, where $f$
is constant for a given shell. 

The Hamiltonian \eqref{eq:inf_h} also matches the gravitational
Hamiltonian (with its negative kinetic term) for a flat Friedmann
model with vanishing 
cosmological constant when
identifying the scale factor as $a\,r_o=R_o$, where $r_o$ is the parametric radius of the dust cloud \cite{OUP}.
When using
Brown-Kucha\v{r} dust as matter and dust proper time as the time
coordinate, the full Hamiltonian constraint for this Friedmann model
reads $H+P_\tau=0$, where $P_\tau$ is the momentum conjugate to $\tau$
\cite{MaedaFriedmann}. Quantizing this 
constraint gives exactly the same Schr\"odinger equation as 
discussed below.  

It follows that all results obtained in the following also apply to
flat Friedmann models with vanishing cosmological constant. The same
holds for models of (marginally bound) Oppenheimer-Snyder collapse, which
shares its dynamics with these cosmological models. 


\section{Quantum dynamics of the outermost shell} \label{ch:chapter_3}

We will now apply the usual canonical quantization procedure in the
Schr\"odinger representation to the Hamiltonian \eqref{eq:inf_h} by
making the substitution 
\begin{equation*}
P_o\to\hat{P}_o=-i\hbar\frac{d}{d R_o}.
\end{equation*}
The operator $\hat{R}_o$ acts by multiplication. In the following
we will suppress the subscript $o$. 
 
The Hamiltonian then reads
\begin{equation}
\hat{H}=\frac{\hbar^2}{2}~R^{-1+a+b}\frac{d}{d
  R}R^{-a}\frac{d}{d R}R^{-b}\label{eq:inf_hq}. 
\end{equation}
The parameters $a$ and $b$ describe our freedom of choosing a factor
ordering. Two possible choices are distinguished. First, $a=b=0$
corresponds to the naive factor ordering in which all derivatives are
on the right. Second, $b=0$ and $a=1/2$ describes the Laplace-Beltrami
ordering, which follows from the demand for covariance in
configuration space. In the following we set $\hbar=1$. 

As a first step towards solving the $\tau$-dependent Schr\"odinger
equation
\begin{equation*}
i\frac{\partial\Psi(R,\tau)}{\partial\tau}=\hat{H}\Psi(R,\tau)
 \end{equation*} 
with the Hamiltonian \eqref{eq:inf_hq}, we derive the stationary modes
$\phi_E(R)$ satisfying $\hat{H}\phi_E=-E\phi_E$, 
\begin{equation}
-E\phi_E=\frac{1}{2}\left(R^{-1+a+b}\frac{d}{d
    R}R^{-a}\frac{d}{d R}R^{-b}
\right)\phi_E \label{eq:inf_hqs}, 
\end{equation}
where $E$ can be interpreted as $E_{\text{ADM}}$.
 
For $E>0$, solutions of \eqref{eq:inf_hqs} are given by
\begin{align}\label{eq:sol_neq0}
\phi^1_E(R)&=R^{\frac{1}{2}(1+a+2b)}~J_{\frac{1}{3}\left| 1+a\right|
             }\!\left(\tfrac{2}{3}\sqrt{2E}R^{\frac{3}{2}} \right),\\ 
\phi^2_E(R)&=R^{\frac{1}{2}(1+a+2b)}~Y_{\frac{1}{3}\left| 1+a\right|
             }\!\left(\tfrac{2}{3}\sqrt{2E}R^{\frac{3}{2}} \right),
      \label{eq:sol_neq0-2}       
\end{align}
where $J_n(z)$ and $Y_n(z)$ are Bessel functions of the first and
second kind, respectively.

The zero energy stationary modes are 
simpler, 
\begin{equation}\label{sol_eq0}
\phi^1_0(R)=R^b\,,\quad \phi^2_0(R)=\begin{cases}
R^{1+a+b}\,,\quad &a\neq-1\\
R^b\,\ln R\,,\quad &a=-1
\end{cases}.
\end{equation}

Although classically $E_{\text{ADM}}\geq 0$, \eqref{eq:inf_hqs} also
possesses solutions for negative energy. They can be interpreted as
genuine quantum solutions without classical counterpart. 
For this case, solutions are 
given by modified Bessel functions $I_n(z)$ and $K_n(z)$, 
\begin{align}\label{eq:sol_-neq0}
\phi^1_{-E}(R)&=R^{\frac{1}{2}(1+a+2b)}~I_{\frac{1}{3}\left|
                1+a\right|
                }\!\left(\tfrac{2}{3}\sqrt{2E}R^{\frac{3}{2}}
                \right),\\ 
\phi^2_{-E}(R)&=R^{\frac{1}{2}(1+a+2b)}~K_{\frac{1}{3}\left|
                1+a\right|
                }\!\left(\tfrac{2}{3}\sqrt{2E}R^{\frac{3}{2}}
                \right). \label{last_mode_standing} 
\end{align}
Note that in the following $E$ will always be positive, and negative
energy stationary states correspond to $-E$. 
We note that for the Laplace-Beltrami
factor ordering, $a=1/2$, the Bessel functions
can be written as elementary functions. 

We will construct the full quantum theory for
our collapsing dust shell in analogy to ordinary quantum
mechanics. We impose square-integrability on wave functions and let
them evolve unitarily according to a self-adjoint Hamiltonian.
This corresponds to enforcing  probability conservation in dust proper
time. The treatment is similar in spirit to the treatment of the
collapsing null dust shells in \cite{HajicekKieferNullShells}.

We start by choosing as the Hilbert space
$L^2(\mathbb{R}^+,R^{1-a-2b}dR)$ the space of square integrable
functions on the positive half-line with respect to the scalar product 
\begin{equation*}
\left\langle\phi,\psi \right\rangle =\int_{0}^{\infty}dR~R^{1-a-2b}\phi^*(R)\psi(R).
\end{equation*} 
The weight $R^{1-a-2b}$ is fixed by the requirement that $\hat{H}$
be symmetric.
\footnote{One can also consider other weights of the form $R^c$ with real
parameters $c$.
Instead of choosing a factor ordering that renders the Hamiltonian
symmetric, equivalently to the above, 
one can construct a symmetrized Hamiltonian of the form
$\frac{1}{2}(\hat{H}+\hat{H}^\dagger)$ (ignoring boundary terms). This
leads to quantum theories equivalent to the one discussed
here, but only if $\min\{1,-a \}\leq b+\frac{c}{2}\leq\max\{1,-a
\}$. If this condition is not fulfilled, additional damping and
potential terms would have to be introduced into the symmetrized Hamiltonian,
or one would have to use complex parameters determining the factor
ordering.}
For Laplace-Beltrami ordering, the weight is just
$\sqrt{R}$. 

We note that we limit our discussion to
stationary solutions of the Schr\"odinger equation and linear
superpositions over different energies constructed from them, with
wave packets in mind. This may exclude some wave functions if
the stationary modes do not form a (generalized) basis of the functions we
are interested in. Whether or not this is the case is hard to
prove rigorously and will not be done here. 
We expect that the wave functions excluded by this restriction, should there be any, are
not physically relevant.   

\subsection{Square integrability}

We will now check which of the stationary modes
\eqref{eq:sol_neq0}--\eqref{last_mode_standing} are square integrable
with respect to our inner product.  
Obviously, the zero energy modes \eqref{sol_eq0} are either not square
integrable at $R=0$ or at $R\to\infty$. The positive energy
modes \eqref{eq:sol_neq0} and \eqref{eq:sol_neq0-2}
are also not square integrable. This can be seen from the
expansion of the Bessel functions for large arguments
\cite{AbramowitzStegun}, 
\begin{align*}
&J_\nu(z)\sim\sqrt{\frac{2}{\pi z}}\cos\!\left( z-\tfrac{1}{2}\nu\pi-\tfrac{1}{4}\pi\right) ,\quad|\arg z|<\pi,\\
&Y_\nu(z)\sim\sqrt{\frac{2}{\pi z}}\sin\!\left(  
                                                                                                                      z-\tfrac{1}{2}\nu\pi-\tfrac{1}{4}\pi\right) 
                                                                                                                    ,\quad|\arg
                                                                                                                    z|<\pi.    
\end{align*}
It follows that the modes $\phi^1_E$ and $\phi^2_E$ approach infinity as
\begin{align}
  R^{\frac{1}{2}(1-a-2b)}\phi^1_E&\sim\frac{R^{\frac{1}{4}}}{\sqrt{\frac{\pi}{3}}(2E)^{\frac{1}{4}}}~\cos\!\left(\tfrac{2}{3}\sqrt{2E}R^{\frac{3}{2}}-\theta_a\right),
  \label{infinity1} \\ 
R^{\frac{1}{2}(1-a-2b)}\phi^2_E&\sim\frac{R^{\frac{1}{4}}}{\sqrt{\frac{\pi}{3}}(2E)^{\frac{1}{4}}}~\sin\!\left(\tfrac{2}{3}\sqrt{2E}R^{\frac{3}{2}}-\theta_a\right)
                                 , \label{infinity2}
\end{align} 
where $\theta_a=\tfrac{\pi}{6}\left| 1+a\right|+\tfrac{\pi}{4}$. We
note that in case of the Laplace-Beltrami factor ordering, this
asymptotic behavior is exact for all $R$.  
That positive energy modes are not square integrable is not
surprising. This is well known from, for example, the case of a free
particle. The solutions are oscillatory and allow an interpretation in
terms of Gel'fand triples (the factor $R^{\frac{1}{4}}$ does not
prevent this). As in quantum mechanics, square integrability can be
achieved by constructing wave packets.

We are now left with the negative energy modes \eqref{eq:sol_-neq0}
and \eqref{last_mode_standing}.
The expansion of the modified Bessel functions for large arguments reads
\cite{AbramowitzStegun}: 
\begin{align}
&I_\nu(z)\sim\frac{e^z}{\sqrt{2\pi z}},\quad|\arg
                z|<\frac{\pi}{2},\label{eq:asymp_psie1}\\ 
&K_\nu(z)\sim\sqrt{\frac{\pi}{2 z}}e^{-z},\quad|\arg
                                                                                              z|<\frac{3\pi}{2}\label{eq:asymp_psie2}. 
\end{align}
We see that the mode $\phi^1_{-E}$ must be discarded
because it diverges exponentially at infinity. As for $\phi^2_{-E}$,
it decreases exponentially at infinity, but we still
have to check its behavior for $R\to 0$. For $z\to 0$ we have for
the Bessel function, 
\begin{equation*}
K_\nu(z)\sim\begin{cases}
\frac{\Gamma(\nu)}{2}\left( \frac{z}{2}\right) ^{-\nu},\quad &\Re(\nu)>0\\
-\ln(z),\quad &\nu=0
\end{cases},
\end{equation*}
hence $\phi^2_{-E}$ approaches the singularity as
\begin{equation}
R^{\frac{1}{2}(1-a-2b)}\phi^2_{-E}\sim\begin{cases}
\frac{\Gamma\left( \frac{1}{3}\left| 1+a\right|\right) }{2 \left(\frac{1}{3}\sqrt{2E} \right)^{\frac{1}{3}\left|1+a \right|}} R^{1-\frac{1}{2}\left| 1+a \right|}, &a\neq-1\\
-R~\ln\!\left( \tfrac{2}{3}\sqrt{2E}R^{\frac{3}{2}}\right), &a=-1
\end{cases}\label{eq:kick_out_K1}. 
\end{equation}
It is thus square integrable also for $R\to 0$ if $\left| 1+a\right| < 3$. Since
it also decays exponentially at infinity,
$\phi^2_{-E}$ is square integrable for these factor orderings. 

\subsection{Self-adjoint extensions of the Hamiltonian}

We now want to find a domain for the Hamiltonian such that it is
self-adjoint. Here, we will only state the results and refer to
Appendix \ref{sec:app_self_adjoint} for details. 
 
For $|1+a|\geq3$, the Hamiltonian is essentially self-adjoint and its
unique domain is equal to what is called its natural domain, consisting
of all square integrable functions $\psi$ such that $\hat{H}\psi$ is
square integrable as well (in addition to some continuity
conditions).  

Additional conditions emerge for $|1+a|<3$. There we have a $U(1)$
family of self-adjoint extensions given by \eqref{b-condition},
\begin{multline}
\left. -(1+e^{i\theta})\,R^{1-|1+a|}~\frac{d}{d
    R}R^{-\frac{1}{2}(1+a-|1+a|+2b)}\psi\right|_{R\to 0}\\\left.=
  i(1-e^{i\theta})\,R^{1+|1+a|}~\frac{d}{d
    R}R^{-\frac{1}{2}(1+a+|1+a|+2b)}\psi\right|_{R\to
  0}\label{eq:condition_a_not_-1} 
\end{multline}
for $a\neq-1$, and by
\begin{multline}
\left. -(1+e^{i\theta})\,R\,\ln^2\!R~\frac{d}{d
    R}\frac{R^{-b}}{\ln R}\psi\right|_{R\to 0}\\\left.=
  i(1-e^{i\theta})\,R~\frac{d}{d
    R}R^{-b}\psi\right|_{R\to 0} \label{eq:condition_a_-1} 
\end{multline}
for $a=-1$. The extensions are parametrized by an angle
$\theta\in[0,2\pi)$.

One might notice that in \eqref{eq:condition_a_not_-1} and
\eqref{eq:condition_a_-1} the powers of $R$ do not match up, and
one might hence suspect that the dimensions could be wrong. In the
construction of self-adjoint extensions for singular operators one has
to insert a dimensionful parameter into the boundary condition we have
given above in order to make the dimensions match. Usually one chooses
for this parameter a relevant scale for the problem at hand, see e.g.
\cite{FulopSelfAdjoint}. The only meaningful scale in our case is the
Planck scale, which in the units chosen here is equal to one,
and as such is not visible in \eqref{eq:condition_a_not_-1} and
\eqref{eq:condition_a_-1}. We could insert an arbitrary dimensionful
parameter into the above expressions, but it would not influence the
results below in any meaningful way. 

The next step is to compute the spectrum of the Hamiltonian and
obtain the generalized eigenbasis. We will refrain from mathematical
rigor and take the usual shortcut of enforcing the boundary
conditions of the self-adjoint extensions,
\eqref{eq:condition_a_not_-1} and \eqref{eq:condition_a_-1}, where
applicable, on our stationary modes $\phi^1_E$, $\phi^2_E$, and
$\phi^2_{-E}$. The second negative energy mode $\phi^1_{-E}$ is
discarded because it 
is not square integrable and can also not be treated by Gel'fand
triples because it is exponentially increasing.

Let us first consider $|1+a|<3$, and start with $\phi^2_{-E}$, see
\eqref{last_mode_standing}, the last stationary mode 
remaining in the Hilbert space. For the case $a\neq-1$, 
\begin{align*}
	R^{1\mp|1+a|}&\frac{d}{d
                       R}R^{-\frac{1}{2}(1+a\mp|1+a|+2b)}\phi^2_{-E}
                       \\&=-\sqrt{2E}R^{\frac{1}{2}\left(  
  3\mp|1+a|\right) }~K_{1\mp\frac{1}{3}\left| 1+a\right|
  }\!\left(\tfrac{2}{3}\sqrt{2E}R^{\frac{3}{2}} \right)\\ 
	&\overset{R\to0}{\sim}-\frac{3}{2}\,\Gamma\!\left(1\mp\tfrac{1}{3}|1+a|
   \right)\left(\tfrac{1}{3}\sqrt{2E}
   \right)^{\pm\tfrac{1}{3}\left|1+a \right|}.  
\end{align*}
Inserting these expressions into \eqref{eq:condition_a_not_-1} shows
that for $\theta\in(\pi,2\pi)$ the stationary mode $\phi^2_{-E}$
evolves unitarily under the time-dependent Schr\"odinger equation
at one specific energy determined by
\begin{equation*}
	\left(\tfrac{1}{3}\sqrt{2E} \right)^{\tfrac{2}{3}\left|1+a
          \right|}=-\tan\tfrac{\theta}{2}~\frac{\Gamma\!\left(1+\tfrac{1}{3}|1+a|
          \right)}{\Gamma\!\left(1-\tfrac{1}{3}|1+a| \right)}. 
      \end{equation*}
This energy corresponds to a bound state. The condition can only be
fulfilled for values of $\theta$ with $\tan\tfrac{\theta}{2}<0$.    
For other values of $\theta$, \eqref{eq:condition_a_not_-1} is
violated for each energy. It remains to check the
case $a=-1$, for which one finds a similar restriction: 
\begin{equation*}
	\ln	\left(\tfrac{2}{3}\sqrt{2E}
        \right)=\tfrac{3}{2}\tan\tfrac{\theta}{2}. 
\end{equation*}
In contrast to $a\neq-1$, this holds for all $\theta\neq\pi$.

Now we turn to the positive energy modes.
Since in contrast to $\phi^2_{-E}$ they are not square integrable, we
will not interpret them as bound states, but identify them with the
continuous part of the spectrum, $E\in\mathbb{R}^+$. As explained in
detail in
Appendix \ref{sec:boundary_conditions_full}, only the linear
combination 
\begin{multline} 
\phi_E(R)=-\tan\tfrac{\theta}{2}\,\frac{\Gamma\!\left(1+\tfrac{1}{3}|1+a|
  \right)}{\Gamma\!\left(1-\tfrac{1}{3}|1+a| \right)}
\left(\tfrac{1}{3}\sqrt{2E} \right)^{-\tfrac{2}{3}\left|1+a
  \right|}\,\phi^1_E\\
-\cos\!\left(\tfrac{\pi}{3}|1+a|\right)\,\phi^1_E
+\sin\!\left(\tfrac{\pi}{3}|1+a|\right)\,\phi^2_E \label{eq:lin_comb_1} 
\end{multline}
for $a\neq-1$ and
\begin{equation}
\phi_E(R)=\left(\tfrac{3}{\pi}\,\tan\tfrac{\theta}{2}-\tfrac{2}{\pi}\,\ln\left( \tfrac{2}{3}\sqrt{2E}\right) \right) \phi^1_E+\phi^2_E \label{eq:lin_comb_2}
\end{equation}
for $a=-1$ fulfill \eqref{eq:condition_a_not_-1} and \eqref{eq:condition_a_-1}, respectively, for all positive energies. We will consider only $\phi_E$ in the construction of wave packets, which we will undertake below. Note that for $\theta=\pi$, where $\tan\frac{\theta}{2}$ diverges, \eqref{eq:lin_comb_1} and \eqref{eq:lin_comb_2} are not valid and have to be substituted for $\phi^1_E$ on its own.

Aside from $\theta=\pi$ there is also the distinguished value $\theta=0$, for
which the mode \eqref{eq:lin_comb_1} takes the particularly simple
form 
\begin{multline}
	\cos\!\left( \tfrac{\pi}{3}|1+a|\right) \,\phi^1_{E}-\sin\!\left( \tfrac{\pi}{3}|1+a|\right) \,\phi^2_{E}\\
	=R^{\frac{1}{2}(1+a+2b)}~J_{-\frac{1}{3}\left| 1+a\right|
        }\!\left(\tfrac{2}{3}\sqrt{2E}R^{\frac{3}{2}} 	 				\right).\label{eq:alternative_mode}
\end{multline}
We note that one cannot construct a mode of this type that fulfills
\eqref{eq:condition_a_not_-1} and \eqref{eq:condition_a_-1} for all
negative energies, since there we only have $\phi^2_{-E}$ at our
disposal. Hence the negative half line is not part of the spectrum of
the Hamiltonian, and negative energies are restricted to those of
stationary bound states. 

Finally we want to mention that for $|1+a|\geq 3$, where the
Hamiltonian is essentially self-adjoint, there are no bound states,
since $\phi^2_{-E}$ is not square integrable, and $\phi^1_E$ is the
only stationary mode that is available
for constructing wave packets, as we
will see in the next subsection. 

Along the same lines we will see that $\phi^2_{-E}$ also has to be
ruled out for constructing wave packets for $|1+a|\geq3$, such that
for all factor orderings only positive energy wave packets exist.  

\subsection{Wave packets and singularity avoidance} \label{sec:wavepackets2}

We want to construct wave packets by superposing stationary modes of
different energies. Without actually calculating the integral involved
in this procedure, we are able to estimate the behavior of these wave packets towards the singularity from the behavior of the stationary modes they are constructed from. 
This is possible because the stationary modes are
well described by power series with terms of the form
$\sqrt{E}^\alpha\cdot R^\beta$ for $R\to0$ \cite{DLMF}. By
integrating this series term by term and assuming that the function
$A$ below
(the wave packet in energy space) is well behaved, it follows that the leading
term in the full wave packet behaves like the leading term of the stationary mode, 
\begin{align}
\Psi(R,\tau)&=\int_0^\infty d\sqrt{E}~ \phi_{
              E}(R)e^{iE\tau}A\!\left(\!\sqrt{E}\right)\\&\sim
  R^\beta~\int_0^\infty d\sqrt{E}~ \sqrt{E}^\alpha
  e^{iE\tau}A\!\left(\!\sqrt{E}\right) \label{eq:transfer_asymptotics}. 
\end{align}

Note that the Bessel function $Y_\nu(z)$ can only be
expressed by a power series as required above when its order $\nu$
is not an integer, which means that we have to exclude these cases. The same holds for $K_\nu(z)$, but this is only marginally relevant here. 

We first consider $\phi^1_E$. For $z\to0$, $J_\nu(z)$ behaves according to
\begin{equation}
J_\nu(z)\sim\frac{1}{\Gamma(\nu+1)}\left( \frac{z}{2}\right) ^\nu,\quad \nu\neq-1,-2,-3,\dots,
\end{equation} 
and hence $\phi^1_E$ approaches the singularity as
\begin{equation}
R^{\frac{1}{2}(1-a-2b)}\phi^1_E\sim\frac{\left(\frac{1}{3}\sqrt{2E} \right)^{\frac{1}{3}|1+a| }}{\Gamma\!\left(1+\frac{1}{3}\left|1+a\right|\right)}~R^{1+\frac{1}{2}\left|1+a\right|}\to0\,. \label{eq:phi1_r0_square2}
\end{equation}
Not only is $\phi^1_E$ square integrable near the singularity, the probability
distribution $R^{1-a-2b}|\Psi|^2$ for the radius $R$, the norm squared of \eqref{eq:phi1_r0_square2}, even vanishes at
$R\to0$. This behavior then also holds for any wave packet constructed
from $\phi^1_E$: For any such wave packet, regardless of the factor
ordering and the specific function $A$, {\em the probability for the
outermost dust shell to be in the classically singular configuration
$R=0$ is zero}. In this sense these wave packets avoid the
singularity. This criterion for singularity avoidance is close to the
DeWitt criterion, cf. \cite{KKP19}. 

As we have seen in the last subsection, we can only use $\phi^1_E$ on
its own as a basis for wave packets when $\theta=\pi$, or, as we will
see shortly, when $|1+a|\geq3$. For other self-adjoint extensions and
factor orderings, we have to consider the linear combination
\eqref{eq:lin_comb_1}, which also includes $\phi^2_E$. Apart from a
prefactor, $Y_\nu(z)$ 
behaves for $z\to0$ as $K_\nu(z)$ does, which means that $\phi^2_E$ behaves
according to \eqref{eq:kick_out_K1} when approaching the
singularity. It thus follows that $\phi^2_E$ (and $\phi^2_{-E}$) must be excluded for the
construction of wave packets for $|1+a|\geq3$, because those wave
packets would not be square integrable when approaching the
singularity; so only $\phi^1_E$ remains. Similarly, for $\phi^2_E$
singularity avoidance occurs along the same lines as for $\phi^1_E$ only
 when $|1+a|<2$. 

We can see that the singularity is always avoided for factor orderings where
$|1+a|\geq3$ or $|1+a|<2$, with the possible exception of $\frac{1}{3}|1+a|\in\mathbb{N}$. We want to emphasize again that this avoidance holds 
independently of the chosen self-adjoint extension and the specific wave packet. Notably, both the naive ($a=b=0$) and the Laplace-Beltrami factor ordering ($b=0$, $a=\frac{1}{2}$) fall into this category of guaranteed singularity avoidance.
The case $\theta=\pi$ should also be highlighted, because there singularity avoidance occurs independently of the factor ordering.

For the cases where we do not have a guaranteed
singularity avoidance, we have instead the guarantee that the
probability distribution for $R$ \emph{does} have support at the
singularity. Thus, depending on the factor ordering and self-adjoint
extension, either the singularity does play a role or it does not; we
cannot influence this by our choice of wave packet.
It should be noted that the remaining stationary mode
\eqref{last_mode_standing} also does not avoid the singularity for
$2\leq|1+a|<3$. Since in addition to being stationary it has a
negative energy, which moreover depends heavily on the factor ordering
and the choice of self-adjoint extension, it can safely be excluded
when discussing gravitational collapse.

To summarize, we see that singularity avoidance is not only possible
but even guaranteed for a wide class of the quantum models 
considered here, and shows a remarkable robustness under many of the
quantization ambiguities. No artificial fine-tuning is required to
achieve this result. 

\subsection{A unitarily evolving wave
  packet} \label{sec:bouncy_bouncy}

To find out how exactly singularity avoidance is facilitated, we want
to construct a positive energy wave packet. We choose the self-adjoint
extension $\theta=\pi$ in order to use $\phi^1_E$ for its
construction for all factor orderings. 

Useful for the construction of non-stationary modes from $\phi^1_E$ is the closure equation (see e.g.\ \cite{ArfkenWeber}, Eq.\ 11.59)
\begin{equation*}
\int_{0}^{\infty}dx~x~J_\nu(ax)J_\nu(bx)=\frac{\delta(a-b)}{a},\text{ for }\nu>-\tfrac{1}{2}.
\end{equation*}
The Bessel functions form an orthogonal set under the scalar product used above. This property also holds in our Hilbert space for the mode $\phi^1_E$,
\begin{equation*}
\int_{0}^{\infty} dR~R^{1-a-2b}~\phi^1_E(R)\phi^1_{\tilde{E}}(R)=\frac{3}{4\sqrt{E}}~\delta\!\left(\! \sqrt{E}-\sqrt{\tilde{E}}\right) .
\end{equation*}
It is more practical to deal with an orthonormal set of modes, hence we rescale $\phi^1_E$ as
\begin{equation*}
\tilde{\phi}^1_E(R)=\frac{2}{\sqrt{3}}E^\frac{1}{4}~R^{\frac{1}{2}(1+a+2b)}~J_{\frac{1}{3}\left| 1+a\right| }\!\left(\tfrac{2}{3}\sqrt{2E}R^{\frac{3}{2}} \right).
\end{equation*}

Our ansatz for constructing wave packets from stationary solutions
reads, as noted before, 
\begin{equation}
\Psi(R,\tau)=\int_0^\infty d\sqrt{E}~ \tilde{\phi}_E(R)~e^{iE\tau}~A\!\left(\!\sqrt{E}\right). 
\end{equation}
For the function $A(\sqrt{E})$ we choose a Poisson-like distribution similar to
the one used in \cite{HajicekKieferNullShells}
for collapsing null shells,
\begin{equation*}
A\!\left(\!\sqrt{E}\right)=\frac{\sqrt{2}\lambda^{\frac{1}{2}(\kappa+1)}}{\sqrt{\Gamma(\kappa+1)}}\sqrt{E}^{\kappa+\frac{1}{2}}e^{-\frac{\lambda}{2}\sqrt{E}^2}, 
\end{equation*}
where $\kappa\geq0$ and $\lambda>0$ are real parameters.
We note that $\kappa$ is
dimensionless and $\lambda$ has the dimension of length. The function
is normalized, 
\begin{equation*}
\int_0^\infty d\sqrt{E}~A^2\!\left(\! \sqrt{E}\right)  =1.
\end{equation*}
The mean (square root of the) energy
and its width are
\begin{align*}
\overline{\sqrt{E}} &= \int_0^\infty d\sqrt{E}~\sqrt{E}~A^2\left( \sqrt{E}\right)=\frac{1}{\sqrt{\lambda}}\frac{\Gamma\!\left( \kappa+\frac{3}{2}\right) }{\Gamma(\kappa+1)}\,,\\
\Delta\sqrt{E}&=\frac{1}{\sqrt{\lambda}}\sqrt{\kappa+1-\frac{\Gamma^2\!\left( \kappa+\frac{3}{2}\right) }{\Gamma^2(\kappa+1)}}.
\end{align*}

Because we have chosen $A\left( \sqrt{E}\right)$ appropriately, there is a
closed form for $\Psi(R,\tau)$ in terms of Kummer's confluent
hypergeometric function $_1F_1(a;b;z)$ (see e.g.\ \cite{Gradshteyn},
Eq.\ 1 in 6.631),  
\begin{widetext}
\begin{multline}
\Psi(R,\tau)=\sqrt{3}\left(\!\frac{\sqrt{2}}{3}\right)^{\frac{1}{3}\left| 1+a\right|+1}\frac{\Gamma\!\left( \frac{1}{6}\left| 1+a\right|+\frac{\kappa}{2}+1\right) }{\sqrt{\Gamma(\kappa+1)}\Gamma\!\left( \frac{1}{3}\left| 1+a\right|+1\right)}~R^{\frac{1}{2}(1+a+|1+a|+2b)}\\
\quad\times\frac{\lambda^{\frac{1}{2}(\kappa+1)}}{(\frac{\lambda}{2}-i\tau)^{
    \frac{1}{6}\left| 1+a\right|+\frac{\kappa}{2}+1}}~_1F_1\!\left(
  \tfrac{1}{6}\left| 1+a\right|+\tfrac{\kappa}{2}+1;
  \tfrac{1}{3}\left| 1+a\right|+1;-\frac{2
    R^3}{9(\frac{\lambda}{2}-i\tau)}\right) \label{eq:general_wavepacket}. 
\end{multline}
\end{widetext}

\begin{figure}[htbp]
	\centering
	\subfloat[$\lambda=2.2$ and $\kappa=9.8$]{\includegraphics[width=\columnwidth]{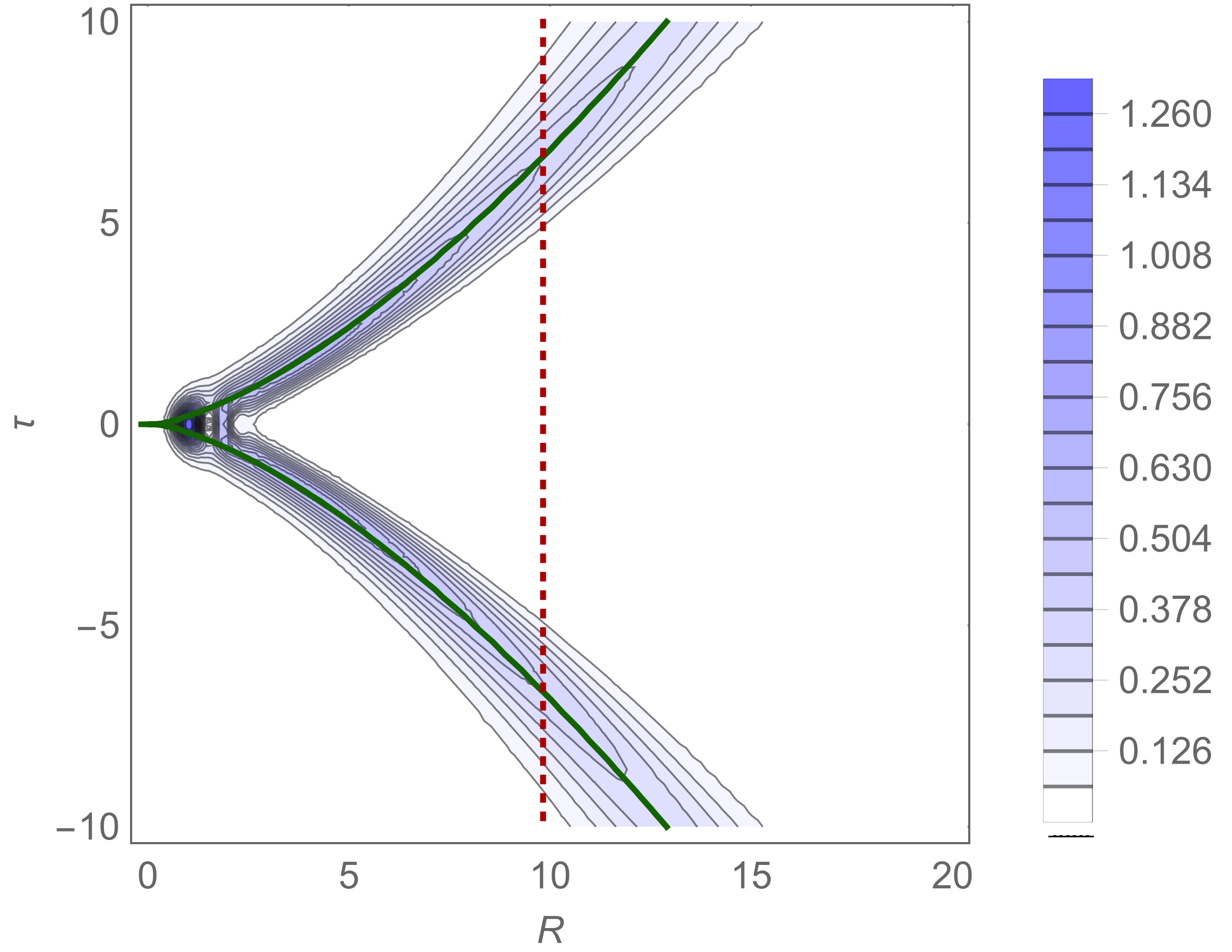} \label{fig:a}}\\
	\subfloat[$\lambda=8.08$ and $\kappa=0.96$]{\includegraphics[width=\columnwidth]{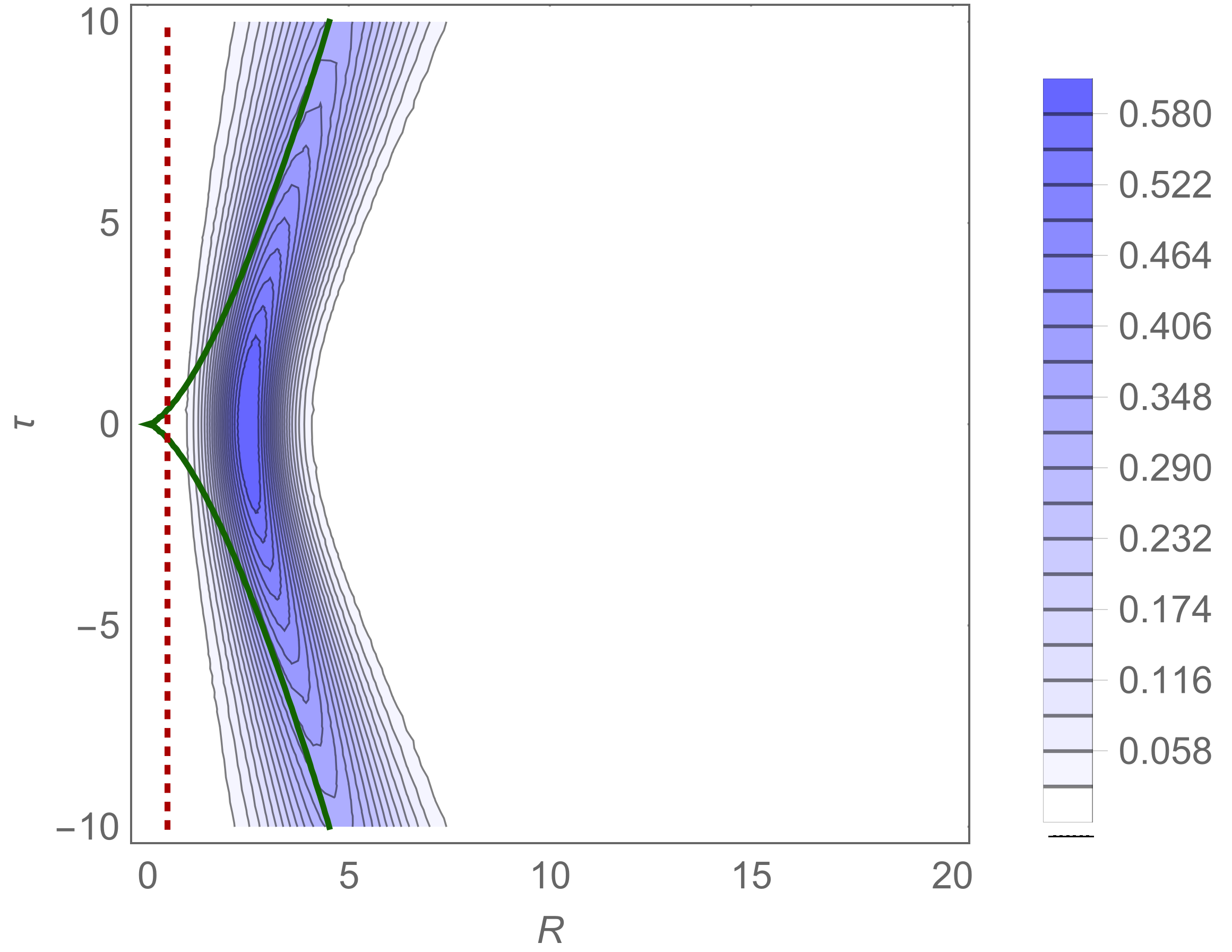}}
	\caption{Probability amplitude for $R$ as given by $R^{1-a-2b}~|\Psi(R,\tau)|^2$, compared to the classical trajectories $R_{cl}=\left(\mp \frac{3}{2}\sqrt{2}\tau\right)^{\frac{2}{3}}\overline{\sqrt{E}^\frac{2}{3}} =\left(\!\mp \frac{3}{2}\sqrt{\frac{2}{\lambda}}\tau\right)^{\frac{2}{3}}\frac{\Gamma\left( \kappa+\frac{4}{3}\right) }{\Gamma\left( \kappa+1\right)}$ (full green line) and the exterior apparent horizon $R_{AH}=2\overline{E}=2\frac{\kappa+1}{\lambda}$ (dotted red line), with $a=2$ and $b=1$, and different $\lambda$ and $\kappa$.}\label{fig:wavepacket_general}
\end{figure}

The behavior of the wave packet can be seen in
Fig.~\ref{fig:wavepacket_general}. It first follows the infalling
classical trajectory up to some minimal $R$ and then makes a transition to
the outgoing classical trajectory: the outermost shell of a collapsing
LTB model bounces before reaching the singularity. Depending on the
parameters of the wave packet, the shell can even fall significantly
far below the apparent horizon until it switches from collapse to
expansion. It should be emphasized that this transition is classically
forbidden and can be interpreted as tunneling from a collapsing to an
expanding configuration, or, in a heuristic picture, from BH to WH. 

So far this model shares its main features with the quantum collapse
of a null shell
\cite{HajicekKieferNullShells,HajicekQuantumNullShells}, but in one
aspect it differs: the wave packet describing the null shell shows
little dispersion, while in our case the wave packet increases in width when
proceeding away from the singularity. This is in contrast to 
minisuperspace models in quantum cosmology, where dispersion near
the singularity was interpreted as a mechanism for singularity
avoidance; see, for example, \cite{KKP19,DKS}. 

We note that the probability distribution for the radius $R$
shows oscillatory behavior near the $\tau=0$ line for high energies,
see Fig.\ \ref{fig:a}. This interference-like pattern emerges
because in this region the part of the wave packet centered around the
classical collapsing trajectory is superposed on the wave packet
around the expanding trajectory. In this sense one could also state
that the singularity avoidance results from destructive interference
between two separate wave packets corresponding to BH and WH,
respectively. 
 
We also note that the general form of
Fig.~\ref{fig:wavepacket_general} does not seem to change with the
factor ordering; the bouncing behavior is always present. In fact, the
parameter $b$ completely cancels out in the probability
distribution $R^{1-a-2b}~|\Psi(R,\tau)|^2$. The details of this
distribution depend, however, on $a$,
such as the position of its peak at $\tau=0$.
 
One can demonstrate that this bouncing behavior shows a certain
robustness also under other details of the quantization: for
$\theta=0$, one can choose the mode \eqref{eq:alternative_mode}
for the construction of wave packets as long as $|1+a|<3$. Due to the
similarity of this mode to $\phi^1_E$ one can extend our wave packet
to this case by simply introducing a few negative signs at places where the
order of the Bessel function enters. Checking the corresponding plots shows
that this wave packet still bounces. For some factor orderings this may even
happen out from a singular configuration. We see that this behavior is
not only robust under changes of the factor ordering, but also under
different choices of self-adjoint extension. 

To discuss the bouncing behavior more rigorously we want to calculate,
for example, the expectation value of the radius $R$ of the outermost
shell. In its current form, our wave packet is too complex to perform
concrete calculations, but fortunately it can be significantly simplified.
We set $\kappa=\frac{1}{3}|1+a|$ and use the identity
$_1F_1(a;a;z)=e^z$ (\cite{DLMF}, eq.\ 13.6.1) to arrive at the wave packet 
\begin{multline}
\tilde{\Psi}(R,\tau)=\sqrt{3}\frac{R^{\frac{1}{2}(1+a+|1+a|+2b)}}{\sqrt{\Gamma\!\left(
      \frac{1}{3}\left| 1+a\right|+1\right)}} \left(
  \frac{\frac{\sqrt{2\lambda}}{3}}{\frac{\lambda}{2}-i\tau}\right)^{\frac{1}{3}\left|
    1+a\right|+1}\\\times \exp\!\left( -\frac{2
    R^3}{9(\frac{\lambda}{2}-i\tau)}\right)  \label{eq:wavepacket_simple}. 
\end{multline}
In quantum cosmology, a similar trick was used in 
\cite{AlvarengaFactorOrdering}. 

As we will see below, by this simplification we have gained the
ability to compute quantities such as $\overline{R}(\tau)$
analytically, but of course this comes at a cost. We can not
independently adjust $\overline{\sqrt{E}}$ and $\Delta\sqrt{E}$
anymore, since both are now proportional to $1/\sqrt{\lambda}$. The
relative width in energy of the wave packet is now fixed by the factor
ordering to 
\begin{align*}
\frac{\Delta\sqrt{E}}{\overline{\sqrt{E}}}&=\sqrt{\frac{\Gamma\!\left(\frac{1}{3}|1+a|+2
                                            \right)\Gamma\!\left(\frac{1}{3}|1+a|+1
                                            \right)
                                            }{\Gamma^2\!\left(\frac{1}{3}|1+a|+\frac{3}{2}
                                            \right)}-1}\\&\leq\left. \frac{\Delta\sqrt{E}}{\overline{\sqrt{E}}}\right|_{a=-1}\approx
  0.53. 
\end{align*}
We see that this wave packet can be rather broadly peaked on its mean
energy, depending on $a$. To decrease its width significantly, one has
to consider factor orderings far beyond the usual ones:
$\frac{\Delta\sqrt{E}}{\overline{\sqrt{E}}}\approx0.2$ for $a=14$, and
$\frac{\Delta\sqrt{E}}{\overline{\sqrt{E}}}\approx0.1$ for $a=71$. As
we have stated above, the bouncing behavior of
\eqref{eq:general_wavepacket} is still present for high values of
$|1+a|$, hence it seems reasonable that results for a $\tilde{\Psi}$
with some well-defined energy (and therefore very high or low $a$)
will also be applicable similarly to more reasonable values of $a$
when considering wave packets of the form
\eqref{eq:general_wavepacket} and narrow in energy. 

For \eqref{eq:wavepacket_simple} we can now compute $\overline{R}$ and $\Delta R$,
\begin{align}
\overline{R} &= \left(\frac{9\lambda}{8}+\frac{9\tau^2}{2\lambda}
               \right)^{\frac{1}{3}} \frac{\Gamma\!\left(
               \frac{1}{3}\left| 1+a\right|+\frac{4}{3}\right)
               }{\Gamma\!\left( \frac{1}{3}\left|
               1+a\right|+1\right)},\label{R-bar} \\
\Delta R&=\overline{R}~\sqrt{\frac{\Gamma\!\left( \frac{1}{3}\left|
          1+a\right|+\frac{5}{3}\right) \Gamma\!\left(
          \frac{1}{3}\left| 1+a\right|+1\right)}{\Gamma^2\!\left(
          \frac{1}{3}\left|
          1+a\right|+\frac{4}{3}\right)}-1}\label{eq:width_radius}\\&\leq\left. \Delta
  R\right|_{a=-1}\approx  0.37\cdot\overline{R}. 
\end{align}
As expected, $\overline{R}(\tau)$ is symmetric in $\tau$ and has a global minimum at $\tau=0$, the minimal radius scaling inversely with the energy for fixed relative width $\frac{\Delta\sqrt{E}}{\overline{\sqrt{E}}}$, 
\begin{align}
R_0&:=\overline{R}(0)=\left(\tfrac{9}{8}\lambda\right)^{\frac{1}{3}}
     \frac{\Gamma\!\left( \frac{1}{3}\left|
     1+a\right|+\frac{4}{3}\right) }{\Gamma\!\left( \frac{1}{3}\left|
     1+a\right|+1\right)}\propto\frac{1}{\overline{E}^{\frac{1}{3}}}
     \label{eq:minimal_radius}.  
\end{align}

That the dependence of $R_0$ on the energy carries over to
\eqref{eq:general_wavepacket} can be checked analytically. One finds that
\begin{equation*}
	\overline{R}(\tau=0)=\lambda^\frac{1}{3}\,g(a,\kappa).
\end{equation*}
The function $g(a,\kappa)$ is rather complicated and can be found in
Appendix \ref{sec:app_R0_full}. When keeping the relative width (and
hence $\kappa$) and the factor ordering constant, this expression is
proportional to $\overline{E}^{\,-\frac{1}{3}}$, as for the simplified
wave packet. Furthermore, it seems that $\overline{R}(\tau=0)$
increases with decreasing relative width in energy and with increasing
$|1+a|$, but a more rigorous analysis is prevented by the complicated
form of $g(a,\kappa)$. 

This result is in contradiction to \cite{RovelliPlanckStars}, in which
by heuristic arguments $R_0\propto E^n$, with $n=\frac{1}{3}$ or
$n=1$, was obtained. Our considerations predict (in the language of
\cite{RovelliPlanckStars}) a Planck star,
meaning a temporary compact remnant of gravitational collapse, with
 sub-Planckian size. For example, for a dust cloud with solar
mass (taking $\kappa=24$, meaning
$\frac{\Delta\sqrt{E}}{\overline{\sqrt{E}}}\approx0.1$ and $a=1$) we
get $\overline{R}(\tau=0)\approx10^{-13}~l_{\rm P}\approx10^{-48}$ m. 

One has to be careful when interpreting this result. Recall that we
only consider the outermost dust shell, but during the bounce the
order of the shells might get reversed, as suggested by the inverse
scaling of $R_0$ with $E$. Remarkably, in that case the size of the
compact object is not necessarily connected to the total mass of the
initial dust cloud, but rather to its structure near the center. The
minimal size of the dust cloud, potentially equal to the minimal
radius of the innermost dust shell, might then be considerably
higher. For example, with the Planck mass as $\overline{E}$ and the
other parameters kept the same, $R_0$ is of the same order of
magnitude as the Planck length. We will present more details on various
aspects of this 
remnant in Sec.~IV and return now to the simplified wave
packet and the corresponding expectation value $\overline{R}$. 

We can show analytically that $\overline{R}(\tau)$ is approximated
very well by classical trajectories when far away from the
singularity, as illustrated in Fig.\ \ref{fig:wavepacket_general}. For
$\tau^2\gg\lambda^2$, 
\begin{align}
\overline{R}(\tau)&\approx\left(\frac{3}{2}\sqrt{2}\left| \tau\right|
                    \right)^\frac{2}{3}\frac{\Gamma\!\left(
                    \frac{1}{3}\left| 1+a\right|+\frac{4}{3}\right)
                    }{\lambda^{\frac{1}{3}}\Gamma\!
                    \left( \frac{1}{3}\left| 1+a\right|+1\right)}\nonumber \\
&=\left(\frac{3}{2}\sqrt{2}\left| \tau\right|                                                                              \right)^\frac{2}{3}\overline{\sqrt{E}^\frac{2}{3}}\label{eq:semicl}.   
\end{align}
It is straightforward to see that this is a solution to \eqref{eq:eom_ltb} for $R(\tau=0)=0$.

\section{Quantum corrected spacetime for dust collapse} \label{ch:chapter_4}
Based on the dynamics of the wave packet discussed in the last section, one can construct a quantum corrected spacetime describing bouncing dust collapse. In this section, we will discuss some aspects of this spacetime.

We take the marginally bound LTB metric,
\begin{equation*}
ds^2=-d\tau^2+(\partial_\rho R)^2~d\rho^2+R^2~d\Omega^2,
\end{equation*}
and use the quantum dynamics of the outermost dust shell to fix the
function $R(\tau,\rho)$. We will focus our discussion on heavy dust
clouds and on corresponding wave packets with a narrow width, such
that they follow the classical trajectories far behind the horizon.  

Depending on what we want to discuss it suffices to simply set
$R(\rho\to\rho_o)=\overline{R}$, such that the trajectory of the
outermost shell matches the expectation value of the corresponding
wave packet. Thereby we leave the evolution of the other shells
completely open, except that they be contained in the outermost
shell at least far away from the singularity. This is the case for our
investigation of the horizon and its lifetime. To compute the
effective pressures arising near the bounce of the quantum corrected
spacetime, we have to make use of the fact that our Hamiltonian gives
the correct dynamics for every single shell, and generalize
$\overline{R}$ to $\overline{R}(\rho)$. 

We will see that at some points further corrections must be made
in order to
account for some inconsistencies of this spacetime. Hence we will
recall the quantum theory in the background and evoke some of its
properties other than the corrected dust trajectories where
necessary. 

\subsection{Horizon}

We have already mentioned that in classical dust clouds apparent
horizons appear where the condition $F(\rho)=R(\tau,\rho)$ is
fulfilled. Attaching a Schwarzschild exterior to the classical LTB
model, an apparent horizon can pass to this exterior from the
outermost shell when the radius of that shell becomes smaller than
$2E_\text{ADM}$. Hence it is the outermost dust shell that determines
the position of this horizon via the mass contained in it, and whether
the horizon is future or past via the sign of its velocity. We will
see in the following that in our quantum corrected spacetime
the exterior horizon's behavior is not quite
as easily determined. 

First we want to determine the position of the horizon. Calculating
the Misner-Sharp mass for the corrected trajectory $R(\tau)=\left(
  R_0^3+\frac{9E}{2}\tau^2\right) ^\frac{1}{3}$, one finds 
\begin{equation*}
M_{\text{MS}}=E~\frac{R^3-R_0^3}{R^3},
\end{equation*}
see \eqref{R-bar} and \eqref{eq:minimal_radius}.
We have seen previously that for heavy dust clouds $R_0\ll2E$, meaning
$R_{\rm AH}=2E$ is still approximately the position of the apparent
horizon in question for early and late times, since $2E\approx
2M_{\text{MS}}(R\gg R_0)$. 

Close to the bounce the situation is more complicated. Because
$M_\text{MS}$ changes in time, one cannot simply match the dust cloud
to a Schwarzschild solution at the outermost shell. As we will see in
Sec.\ \ref{sec:effective_matter}, effective pressures occur in our
quantum corrected spacetime, which further prevent the matching to an
exterior region \cite{Joshi}. Taking the exterior apparent
horizon to be at $R_{\rm AH}=2M_{\text{MS}}$, we can see that when
approaching the bounce the apparent horizon shrinks and even
disappears for $R=R_0$, which means that it
will vanish back into the dust cloud for some time during the bounce.
This is in agreement with
other propositions for the behavior of the horizon in similar models
\cite{MalafarinaBounceRev}. 

In the following, we will assume that an exterior horizon is
present at $R=2E$ as long as the outermost shell is inside this
radius, since this introduces the least radical modification into the
corrected spacetime. This leaves us to explain the transition of the
horizon from BH to WH. 

Recall that whether the horizon in question is future or past is
determined by the sign of $\dot{R}$. For the BH it is negative,
while it is positive for the WH. If we limit ourselves to just the
quantum corrected spacetime, the horizon will of course be either
future or past, with an instantaneous transition when the shell turns
around. To smooth out this process we can invoke the quantum model and
allow the horizon to be in a superposition, as was done in
\cite{HajicekQuantumNullShells}. 

Classically, the momentum $P=-2R\dot{R}$ always has the opposite sign to $\dot{R}$, meaning the nature of the horizon can be determined with the help of the operator $\hat{P}$. Unfortunately, as is well known for the momentum operator on the half-line, it cannot be made self-adjoint, meaning $\hat{P}$ is not technically an observable. Nevertheless, for the calculation of an expectation value a symmetrized version of $\hat{P}$ is sufficient. The operator
\begin{equation*}
\hat{\mathsf{P}}=-i~R^{-\frac{1}{2}(1-a-2b)}\frac{\partial}{\partial R}R^{\frac{1}{2}(1-a-2b)}
\end{equation*}
fits our purposes. Now we can calculate the expectation value of
$\hat{\mathsf{P}}$ with respect to the simplified wave packet \eqref{eq:wavepacket_simple},
\begin{align*}
\overline{\mathsf{P}}&=-i~\left( 1+\tfrac{1}{2}\left|1+a \right| \right)~\overline{R^{-1}}+i~\frac{2}{3\left( \frac{\lambda}{2}-i\tau\right) }\overline{R^2}\\
&=-3\tau~\left(\frac{9\lambda}{8}+\frac{9\tau^2}{2\lambda} \right)^{-\frac{1}{3}}\frac{\Gamma\!\left( \frac{1}{3}\left| 1+a\right|+\frac{5}{3}\right) }{\lambda\Gamma\!\left( \frac{1}{3}\left| 1+a\right|+1\right)}\propto-\tau.
\end{align*}
This shows the behavior one would expect: before the bounce we have
$\sgn \overline{\mathsf{P}}>0$ and hence a BH horizon, and afterwards
with $\sgn \overline{\mathsf{P}}<0$ a WH horizon. We can make an educated guess concerning the transition in between by normalizing $\overline{\mathsf{P}}$ by the condition that at $\tau\to-\infty$ the wave packet was in a pure BH state, to which we assign the value $1$ (and correspondingly to a WH $-1$), leading to
\begin{equation}
\overline{\mathsf{p}}=\frac{\overline{\mathsf{P}}}{\overline{\mathsf{P}}_{\tau\to-\infty}}=-\sgn\tau~\left( \frac{\tau^2}{\frac{\lambda^2}{4}+\tau^2}\right)^{\frac{1}{3}}\label{eq:black_hole_ness}\,.
\end{equation}
$\overline{\mathsf{P}}_{\tau\to-\infty}\propto\left(\frac{2}{9}\lambda|\tau| \right)^{\frac{1}{3}} $ is the asymptotic behavior of $\overline{\mathsf{P}}$ at very early times, for the normalization extended to all $\tau$. Taking $\overline{\mathsf{p}}$ as a measure of `black hole-ness', we see that the transition from BH to WH is instantaneous for $\lambda\to0$, and smoothed out for higher values of the parameter. Note that the minimal radius \eqref{eq:minimal_radius} scales with a positive power of $\lambda$. It follows that the closer the wave packet comes to the singularity, the more rapid is the transition of the horizon.

Taking into account that during the bounce the order of the shells
might get `scrambled' such that the outermost shell need not stay outermost, it would be appropriate to alter the exact form of the horizon transition to reflect the behavior of the shell that actually has the largest radius at a given $\tau$. We would then expect a further smoothing of the transition.

\subsection{Lifetime}\label{sec:lifetime}

The lifetime of the exterior horizon is of great interest as a
consistency check of our model. It should be long enough such that the
bouncing collapse at least resembles a BH; otherwise, this scenario would
be excluded by astrophysical observations. 

In order to discuss this lifetime we introduce two observers into the
spacetime, one at a fixed physical radius $R_{\rm obs}$ and the other
comoving with the dust cloud. These two observers will meet twice,
first during the collapse and again during the re-expansion. The time
difference between these two events for the comoving observer is then given by 
\begin{align*}
&\Delta\tau=\tau_+-\tau_-=\sqrt{\frac{8R_{\rm obs}^3}{9}\frac{\lambda\Gamma^3\!\left(\frac{1}{3}\left|1+a \right|+1\right) }{\Gamma^3\!\left(\frac{1}{3}\left|1+a \right|+\frac{4}{3}\right)}-\lambda^2},
\end{align*}
where $\tau_\pm$ is defined by $R_{\rm obs}=\overline{R}(\tau_\pm)$. For a
heavy cloud and a fixed relative width in energy, $\lambda$ has to be
small; we can thus neglect the second term under the square root
and find 
\begin{align*}
\Delta\tau&=\sqrt{\frac{8R_{\rm obs}^3}{9}}\left(
            \frac{\lambda^\frac{1}{3}\Gamma\!\left(\frac{1}{3}\left|1+a
            \right|+1\right) }{\Gamma\!\left(\frac{1}{3}\left|1+a
            \right|+\frac{4}{3}\right)}\right)^\frac{3}{2}\\&=
  \sqrt{\frac{8R_{\rm
  obs}^3}{9\overline{E^\frac{1}{3}}^3}}\leq\sqrt{\frac{8R_{\rm
  obs}^3}{9\overline{E}}}. 
\end{align*}
The last step follows from H\"older's inequality,
$\overline{X^q}\leq\overline{X^p}^\frac{q}{p}$ for $0<q<p$. For narrow
wave packets one would expect the last two terms to be nearly
equal. This result is equal to twice the free fall time of the
outermost shell from an initial radius $R_{\rm obs}$ down to $R=0$.

The lifetime of the grey hole can then be taken to be $\Delta\tau$
with $R_{\rm obs}=\overline{R_{\rm AH}}$,
\begin{equation*}
\Delta\tau_{\rm GH}\approx\frac{8}{3}~\overline{E}.
\end{equation*}
The lifetime from the point of view of the comoving observer scales linearly with the dust cloud's mass, an unsurprising result given how closely $\overline{R}$ sticks to the classical trajectories. More interesting for comparison with observations is the timescale experienced by the other, external observer.

It is at this point that we run into a problem: The exterior of our
bouncing dust cloud at least at early and late times can be described
via a Schwarzschild black hole or, more precisely, appropriate patches
of the Kruskal spacetime. In terms of Schwarzschild Killing time, which
a stationary observer very far from the dust cloud approximately
experiences, crossing the apparent horizon (which for a heavy dust
cloud happens at sufficiently early and late times) takes infinitely
long. This prediction seems paradoxical: The comoving observer returns
in finite time to his exterior counterpart, for whom an infinite
amount of time has passed. The outside observer would see his more
adventureous friend as being stuck when approaching the apparent horizon.

It appears that further modification of the quantum corrected
spacetime is necessary, as was also argued in
\cite{HaggardRovelliBounce,BarceloBounce1}, and in a different
context in \cite{IsraelBounce}. Unfortunately, our model is formulated
in terms of dust proper time, and we have cut off the exterior
geometry. Hence calculating the lifetime as seen from the exterior
observer would entail transforming to Schwarzschild Killing time,
which is ill-defined in the quantum model, since this transformation
depends on $R$ and on the energy $E$. Attaching an exterior to the
dust cloud is also problematic, as we have discussed in the last
section; it is also ambiguous
because the time delay between horizon crossings is an open parameter
\cite{HaggardRovelliBounce}.  

We will instead follow a different approach by incorporating 
another physical mechanism into the quantum corrected spacetime
picture: transitions between dynamically distinct `states' of the dust
cloud, sticking closely to the picture of BHWH tunneling as employed
in \cite{ChristodoulouLifetime2,ChristodoulouLifetime}. There, the
lifetime of a bouncing null dust shell was computed in a way
which in the following we will adapt to our model. 

We differentiate between three states of the dust cloud:
collapsing while being at least partially outside its horizon; being
completely inside the horizon (referred to below as the grey hole);
and expanding outside of its horizon.

We then consider the following setup. The cloud, characterized by its
outermost shell, behaves semiclassically up until close to the horizon,
in accordance with our previous results. Due to the
aforementioned gravitational time dilation, quantum-gravitational
effects have a chance to accumulate. At this point, the dust cloud will
inevitably experience a transition to one of the other states listed
above. Furthermore, motivated by results for the BHWH tunneling timescale
\cite{ChristodoulouLifetime,BarceloLifetime,AmbrusHajicekLifetime},
we assume that the transition itself takes a
relatively short amount of time, roughly proportional to the mass of
the dust cloud.

This accumulation, or pile-up, of quantum effects when approaching the
horizon was first proposed by Haggard and Rovelli in
\cite{HaggardRovelliBounce}. We want to note that this mechanism cannot
straightforwardly be applied as an explanation for the transition
of the horizon, as there one cannot take the distinguished notion of time to be Schwarzschild Killing time.

To determine the lifetime we need to compute the relevant transition probabilities. We will take these probabilities to be determined by our quantum model,
\begin{align*}
W(\tau_-,\tau_+)&=\left| \int_{0}^{\infty} dR~R^{1-a-2b}~\tilde{\Psi}^*(R,\tau_-)~\tilde{\Psi}(R,\tau_+)\right| ^2\\&=\left(\frac{\lambda^2}{\lambda^2+(\tau_+-\tau_-)^2}\right)^{\frac{1}{3}\left|1+a\right|+1 }.
\end{align*}

The three states can then be characterized by ranges in proper time:
$\tau<-\tau_{\rm AH}$ for collapse (C), $-\tau_{\rm AH}<\tau<\tau_{\rm
  AH}$ for the grey hole (GH), and $\tau_{\rm AH}<\tau$ for
expansion (E). $\pm\tau_{\rm AH}$ with $\tau_{\rm AH}>0$ are the proper
times at which the outermost shell reaches the apparent horizon,
$\overline{R}(\pm\tau_{\rm AH})=2\overline{E}$. 

Let us now follow the dust cloud from the collapsing to the expanding
state. First, the outermost shell approaches the apparent horizon from
the outside and will eventually make a transition either to the grey
hole or to the expanding state. Which case is more likely?  

To answer this question, let us consider the transition probabilities
\begin{align*}
\frac{P_{{\rm C}\to {\rm E}}}{P_{{\rm C}\to {\rm GH}}}&=\frac{\int_{-\infty}^{-\tau_{\rm
                                AH}}d\tau_-\int_{\tau_{\rm
                                AH}}^{\infty}d\tau_+~W(\tau_-,\tau_+)}{\int_{-\infty}^{-\tau_{\rm 
                                AH}}d\tau_-\int_{-\tau_{\rm
                                AH}}^{\tau_{\rm
                                AH}}d\tau_+~W(\tau_-,\tau_+)}\\&\approx\frac{\left(2\frac{\tau_{\rm 
  AH}}{\lambda}\right)^{-\frac{2}{3}\left|1+a \right|}
  }{\frac{2}{3}\left|1+a \right|+1}. 
\end{align*}
This is an approximation for high energies of the full expression,
which can be found in Appendix \ref{sec:app_lifetime}. We have used
that ${\tau_{\rm AH}}/{\lambda}$ roughly scales with
$\overline{E}^2$. It follows that for high energies (and non-maximal
relative widths of the wave packet, $a\neq-1$) the transition to the
grey hole state dominates. As a result we will focus on this
transition. 

We will now define the lifetime as the time it takes for the dust cloud
to make a transition from grey hole to the expanding state. Once in this
state, the outermost shell will expand away from the apparent horizon
and will not get the chance to make a transition to a different state
again. It will stay outside its horizon, and the grey hole is gone. To
determine this lifetime we follow \cite{ChristodoulouLifetime},
where a BH lifetime was computed using a picture of BHWH tunneling,
and draw an analogy to an alpha particle tunneling out of a nucleus.
A simple model for this process is the following: the particle
travels across the nucleus and after a time $\Delta t$ hits a potential
wall which it can traverse with probability $p$. If it fails, it will be
reflected and can try again when, after the time $\Delta t$ has
elapsed once more, it hits a potential wall on the other side. The
lifetime of the nucleus can then be estimated as ${\Delta t}/{p}$. 

Taking also the previously discussed transition from collapsing- to GH
state into account, our picture of dust collapse from the perspective
of an exterior observer seems to resemble the quantum mechanical
process of `resonant tunneling', where at specific energies depending
on the potential barrier metastable states can occur during
scattering. Some of the different notions of tunneling time (see e.g.\
\cite{HaugeTunneling}) can also be applied to resonant tunneling (see
e.g.\ \cite{OlkhovskyResonant}). Unfortunately, this requires knowledge
of the full wave function, which we do not possess. So we return to
our picture of three distinct states.

What we need to determine now is the probability for the dust cloud to
evolve from grey hole to expanding state, what process replaces
reflection in the analogy above, and the time it takes until the cloud
is ready to try escaping again. The probability can be determined as
above, 
\begin{align}
P_{{\rm GH}\to {\rm E}}&=\frac{\int_{-\tau_{\rm AH}}^{\tau_{\rm AH}}d\tau_-\int_{\tau_{\rm AH}}^{\infty}d\tau_+~W(\tau_-,\tau_+)}{\int_{-\tau_{\rm AH}}^{\tau_{\rm AH}}d\tau_-\int_{-\infty}^{\infty}d\tau_+~W(\tau_-,\tau_+)}\\
&\approx\frac{ \Gamma\! \left(\tfrac{1}{3}\left| a+1\right| \right)}{4
                                                                                                                                                                                                                                   \sqrt{\pi
                                                                                                                                                                                                                                   }\,
                                                                                                                                                                                                                                   \Gamma\!
                                                                                                                                                                                                                                   \left(\tfrac{1}{3}\left|
                                                                                                                                                                                                                                   a+1\right| 
                                                                                                                                                                                                                                   +\frac{1}{2}\right)}\frac{\lambda}{\tau_{\rm
                                                                                                                                                                                                                                   AH}}. \label{eq:escape_prob}   
\end{align} 
The last line is once again an approximation for high energies. We see
that the probability for the dust cloud to escape the grey hole state
is proportional to ${1}/{\overline{E}^2}$ for heavy clouds. 

Note that the above only holds for $a\neq-1$. For $a=-1$, the escape
probability $P_{{\rm GH}\to {\rm E}}$ behaves to leading order like
$\frac{\lambda}{\tau_{\rm AH}}\,\ln\!\left(2\frac{\tau_{\rm
      AH}}{\lambda}\right) $. This is of no further concern here since
we are only interested in narrow wave packets, but serves as a warning
that for the full wave packet, where the width is not related to $a$,
this result might change for this specific class of factor orderings. 

Classically the only timescale at our disposal is $\overline{E}$, and
hence it seems reasonable to assume that the time between escape
attempts is proportional to $\overline{E}$, as also argued in
\cite{ChristodoulouLifetime}. We will refrain from guessing the
corresponding alternative process at this point and instead leave its
discussion for future work. It might not even be relevant from the
point of view of the exterior observer, because there is the
possibility that whatever happens is hidden behind a horizon. 

Combining our results thus gives a total lifetime proportional to
$\overline{E}^3$. Other contributions are negligible in
comparison: We have assumed the time for the transition itself to be
of the order $\overline{E}$. Furthermore, the Killing time when
approaching the horizon only diverges logarithmically, and hence the
time it takes until the initial transition into the grey hole takes
place is, depending on how close to the horizon this occurs, most
likely appreciably smaller than $\propto\overline{E}^3$. 

Our lifetime is considerably larger than earlier results that predict
a lifetime linear in
$\overline{E}$ \cite{AmbrusHajicekLifetime,BarceloLifetime}. It has
since been argued that this describes only the time for the transition
itself, in case this happens, and should be complemented by a
timescale associated with the failure to perform a transition
\cite{ChristodoulouLifetime}. This is also the viewpoint we adopt
here, but compared to the lifetime
$\propto\overline{E}~e^{\Xi\overline{E}^2}$ found in
\cite{ChristodoulouLifetime}, our lifetime is significantly smaller. It
is, in fact, comparable with the Hawking evaporation time, making Hawking
radiation a significant factor for the lifetime. The explicit
inclusion of it deserves further investigation. Also of the same
timescale is the dispersion time of a wave packet describing a
quantized extremal Reissner--Nordstr\"om black hole
\cite{KieferExtremalBHQuant}. 

Our result also further corroborates the usual sentiment that the
semiclassical description of quantum black holes breaks down within a
timescale of $\overline{E}^3$, an idea first introduced in the
discussion of Hawking evaporation and supported by the 
results of \cite{KieferExtremalBHQuant}. In our model, the exterior observer first
notices the bounce when this time has elapsed after the
formation of the grey hole, breaking at least the global notion of a
correct classical description of the geometry far away from the
singularity.

In spite of its limitations, we are confident that our simple model provides
convincing arguments for a finite, but not too short lifetime for
the transition from BH to WH.

\subsection{Effective pressure}\label{sec:effective_matter}

For the following discussion it is necessary to generalize our results
from the outermost dust shell to the full LTB model; we thus assume 
\begin{align*}
&R(\tau,\rho)=\left(R_0(\rho)^3+R_{\rm cl}(\tau,\rho)^3\right)^\frac{1}{3},\\&\text{with}\quad
  R_0(\rho)=\frac{\alpha(\rho)}{F(\rho)^\frac{1}{3}}\quad,\quad\dot{R}_{\rm
  cl}(\tau,\rho)^2=\frac{F(\rho)}{R_{\rm cl}(\tau,\rho)}.   
\end{align*}
Here, $\alpha(\rho)$ has been heuristically introduced to describe
this generalization.
We leave this function open, except for the condition that
when approaching $\rho\to0$, $\alpha\propto F^\frac{1}{3}$ such that the
minimal radius of the innermost shell does not diverge. Furthermore,
$\alpha$ has to be chosen in such a way that for every shell the
initial radius is at least as big as its minimal radius. 

One should note that a special case of this class of bouncing dust
collapse models was discussed in \cite{MalafarinaBounce}, motivated by
a specific correction of the energy density through quantum
effects; the authors of \cite{MalafarinaBounce} considered homogeneous
dust and used a specific function $\alpha$.  

Inserting the resulting metric into the Einstein equations, we can determine an effective energy-momentum tensor. It is diagonal, and hence we interpret its components as an effective energy density and three components of (anisotropic) pressure,
\begin{align*}
8\pi\epsilon&=\frac{1}{R'R^2}\left(F-\frac{\alpha^3}{R^3}\right)',\\
8\pi p_\rho&=-3\frac{\alpha^3}{R^6},\\
8\pi p_\theta&=8\pi p_\phi=\frac{3}{2}\frac{\alpha^3}{R^6}
               -\frac{3}{2}\frac{1}{R'R^2}\left(\frac{\alpha^3}{R^3}\right)'. 
\end{align*}
As we can see, the corrections to the energy density and the pressures
build up quickly very close to the bounce because of the factors $R^{-6}$. To facilitate the bounce, the pressure and the
correction to the energy density need to become negative enough to
make gravity repulsive. Adding up all contributions gives 
\begin{equation*}
8\pi\left( \epsilon+p_\rho+p_\theta+p_\phi\right)
=\frac{1}{R'R^2}\left(F-4\frac{\alpha^3}{R^3}\right)'. 
\end{equation*}
For simplicity we will in the following consider this expression at
the time of the bounce for individual shells. After all, one would
expect that the repulsion is strongest then. This gives
\begin{equation}
\left. 8\pi\left( \epsilon+p_\rho+p_\theta+p_\phi\right) \right|_{R=R_0} =-3\frac{F'}{R_0'R_0^2}\label{eq:repulsive_pressure}.
\end{equation}
It should be noted that in agreement with the Misner-Sharp  mass,
$\epsilon$ vanishes at the bounce. 

The expression \eqref{eq:repulsive_pressure} need not necessarily be
negative.\footnote{It should be noted that the relevant energy
  conditions are still violated, so the singularity theorems
  (see e.g. \cite{HawkingEllis}) are not applicable, allowing the
  possibility of a bounce.
  Consider e.g.\ $\left. 8\pi( \epsilon+p_\rho)
  \right|_{R=R_0}=-3\frac{F}{R_0^3}<0$, which violates the null, weak,
  dominant and strong energy condition.} The shells may get
`scrambled', that is, their order may (perhaps partially) be reversed. Because this can
only be the case for the future of a shell-crossing singularity, one
has to specify how the spacetime is extended through it. We have done
that already in the form of the shell trajectories $R(\tau,\rho)$
above, but have to be aware that the interpretation of some quantities
changes. Most relevant here is the fact that the mass function $F(\rho)$ is
still constant in time, but cannot be equal to the mass contained
in the shell $\rho$ after crossing another shell. We have to interpret
$F$ as a label attached to the shells. This restriction is then
lifted after the shell crossings occur a second time, and $F$ once
again regains its former status. Furthermore, we have to consider that
the coordinates do not have to retain their physical meaning during
the bounce: $\tau$ is not necessarily the dust proper time, a fact
which will not restrict the following considerations, whereas the fact that
$\rho$ is not monotonically increasing when going outwards will become very
important. 

It follows that while $F'$ is always positive, there is no guarantee
that $R'$ will stay positive during the bounce, potentially changing
the sign of the effective energy density
\eqref{eq:repulsive_pressure}. If this happens, gravity can become
repulsive. How can we understand this?

To answer this question we calculate the active gravitating mass
inside the shell $\rho$ by the following integral, at first without
any scrambling, 
\begin{align*}
M(\rho,\tau)&=4\pi\int_{0}^{\rho}d\tilde{\rho}~\sqrt{-g}~\left( \epsilon+p_\rho+p_\theta+p_\phi\right)\\&=\frac{F}{2}-2\frac{\alpha^3}{R^3}\overset{R=R_0}{=}-\frac{3}{2}F\,.
\end{align*}
As expected, gravity becomes repulsive, and also stronger by a factor
of three as compared to the classical collapse.  

We now address the case of scrambling which we restrict
to the case where the order of all shells is completely
reversed. Taking into account that the innermost shell is then the
former outermost one with $\rho=\rho_o$, we have
\begin{align*}
M(\rho,\tau)&=4\pi\int_{\rho_o}^{\rho}d\tilde{\rho}~\sqrt{-g}~\left(
              \epsilon+p_\rho+p_\theta+p_\phi\right)
              \\&=\left[
  -\frac{F}{2}+2\frac{\alpha^3}{R^3}\right]_{\rho_o}^\rho\!\!\!\!
  \overset{R=R_0}{=}\tfrac{3}{2}F(\rho)-\tfrac{3}{2}F(\rho_o)<0. 
\end{align*}
The sign change in the second line is a result of $|R'|$ appearing in
the square root of the metric determinant and of ${R'}^{-1}$ appearing
in the effective energy density and pressure. It is apparent that now
gravity is repulsive not as a result of negative effective pressure,
but simply because the shells are scrambled.

\section{Conclusions}

In this paper, we have quantized the LTB model using the assumption
that the quantum dynamics of
different dust shells decouple, just as in the classical case. This
has allowed us to quantize only a 
single one of those shells, chosen to be the outermost one, and infer
the behavior 
of the full dust cloud from the results. 

Because the dust brings with it a natural time coordinate, its proper
time, we have been able to ignore the usual problem of time in quantum
gravity \cite{OUP}.  This has enabled us to construct a quantum theory
for the outermost shell in analogy to conventional quantum mechanics,
including unitary evolution of states. Both the
choice of factor ordering and self-adjoint extension have been left
open. 

We have been able to show that unitarily evolving states generically
avoid the classical singularity, except when the factor ordering falls
into a specific range. Outside of this range, singularity avoidance
holds for all self-adjoint extensions. Choosing a convenient
self-adjoint extension has allowed us to examine a particular
singularity-avoiding wave packet for all factor orderings. This wave
packet exhibits a bounce. We have demonstrated that this bouncing behavior
exhibits a robustness under quantization
ambiguities similarly to singularity avoidance. 

We have then investigated several properties of a quantum corrected
model for gravitational collapse based on the dynamics predicted by
our quantum theory: the transformation of the horizon 
from black hole to 
white hole, the lifetime of the grey hole, which turns out
proportional to the third power of the ADM energy,
and effective pressures facilitating the
bounce. Regarding the last point, we have found that these pressures
are not negative enough to make gravity repulsive in those cases where
the different dust shells change their order during the bounce, but
there the effective mass inside each shell is still negative exactly
because of this reversed order. 

When discussing these aspects of bouncing collapse, the limits of
applicability of our model became apparent: using
dust proper time as the time parameter and cutting off the model at
the dust cloud's outermost shell has led to difficulties in
determining the grey hole's lifetime and to limitations
in understanding the apparent horizon.  

The perhaps strongest limitation of our model
is the assumption that the shells can be treated
independently from each other.
It is far from clear whether the shells do
or do not show some emergent interaction when quantizing the full LTB
model. In fact, we expect additional terms to occur in the exact Hamiltonian;
after all, Hawking radiation is not accounted for in our model. 
Including this radiation may by itself modify some of our results,
especially in view of the lifetime we have computed (which is of the
same order as the evaporation time). Perhaps it will be possible to
accommodate such effects in an extended model similar to \cite{KMM09}.   

In addition, the possible occurrence of shell crossings near the bounce leaves
some open questions. We have proposed a particular method to deal with
them, but there might be a more elegant alternative which will also
be applicable to the classical shell crossings that we have excluded
from the beginning.

In spite of these limitations, we believe that our results are a first indication
that quantum-gravitational effects can indeed lead to singularity
avoidance in the LTB model, and that the underlying mechanism
is a bounce. The degree of robustness of these features under
the quantization ambiguities is certainly
encouraging. 


\section*{Acknowledgments}

TS thanks the Bonn-Cologne Graduate School (BCGS) for Physics and Astronomy
for financial support, and Yi-Fan Wang, Nick Kwidzinski, and Jens Boos
for helpful discussions. 

\appendix
\section{}\label{sec:app_self_adjoint}

This Appendix is devoted to finding
the self-adjoint extension of the Hamiltonian \eqref{eq:inf_hq}; in this, we
largely follow \cite{GitmanSelfAdjoint}.  To start with, we choose as
the domain of $\hat{H}$ all functions in
$L^2(\mathbb{R}^+,R^{1-a-2b}dR)$ that are smooth and compactly
supported on the half-line such that the boundary term 
\begin{align}
	W(\psi,\phi)&=\left\langle\phi,\hat{H}\psi \right\rangle-\left\langle\hat{H}\phi,\psi \right\rangle\nonumber\\
	&=\left. R^{-a-2b}\left(\phi^*\frac{\partial\psi}{\partial
   R}-\frac{\partial\phi^*}{\partial R}\psi \right)
   \right|_0^\infty,\label{eq:inf_symm} 
\end{align}
where we take $\hat{H}$ just as a differential operator without a
well-defined domain, vanishes for such a function $\psi$,
independently of $\phi\in L^2(\mathbb{R}^+,R^{1-a-2b}dR)$. Hence the
domain of its adjoint is as large as it can be for a second order
differential operator, what is called in \cite{GitmanSelfAdjoint} its
natural domain. 

To find out whether the domain of the self-adjoint Hamiltonian is
unique, we need to find the deficiency indices of $\hat{H}$ as the
dimensions of the solution spaces to the eigenvalue equations
$\hat{H}^\dagger\psi=\pm i\psi$. The corresponding solutions are the
same as the positive and negative energy stationary modes from the
beginning of Sec.~III; one simply has to replace $E$ by $i$. 

Checking for square integrability can also be done in analogy to the
stationary modes: for the eigenvalue $-i$ only one mode remains and
only for factor orderings $\left| 1+a\right| < 3$. Hence we have for
these factor orderings the deficiency index $n_-=1$, and otherwise
$n_-=0$. Because $\hat{H}$ is real, the same has to hold for
$n_+$. Why we have a square integrable solution to the eigenvalue
equation for eigenvalue $i$, but none for a real eigenvalue, can be
seen in the following way. The asymptotic behavior of $\phi_E^{1/2}$
for $R\to\infty$, \eqref{infinity1} and \eqref{infinity2},
acquires an exponential component in addition to an oscillating one
for $E=i$. For a specific combination of the two modes the
exponentially growing parts can be made to cancel out, leaving an
exponential decay towards infinity. 

The deficiency indices tell us that $\hat{H}$ is essentially self-adjoint
for $\left| 1+a\right| \geq 3$, meaning it has a unique 
self-adjoint extension for those factor orderings. For $\left| 1+a\right| <
3$ the extension is not unique, but several choices are possible. Let
us start with the former case. 

The unique self-adjoint extension of an essentially self-adjoint
operator is equal to its closure. The domain of this closure is given
by all functions $\phi \in \text{dom}\,\hat{H}^\dagger$ such that
$W(\psi,\phi)=0$ for all $\psi \in \text{dom}\,\hat{H}^\dagger$. Let
us first note that for every such $\psi$ one can construct a function
such that it and its derivative behave like the original function $\psi$
(or respectively its derivative) at $R\to\infty$ or $R\to0$, and
vanish for the other boundary. It follows that we can split up the
above condition $W(\psi,\phi)=0$ into 
\begin{align}
	&\left. w(\psi,\phi)\right|_{R\to0}=0 \quad \text{and} \quad \left. w(\psi,\phi)\right|_{R\to\infty}=0\,,\nonumber\\
	&\text{where}\quad w(\psi,\phi)=\tfrac{1}{2}\,R^{-a-2b}\left(\phi^*\frac{d\psi}{d R}-\frac{d\phi^*}{d R}\psi \right). \label{eq:w}
\end{align}

To arrive at generic boundary conditions for unitarily evolving wave functions, we have to determine how a generic $\psi\in\text{dom}\,\hat{H}^\dagger$ behaves when approaching the boundaries. Let us first consider $R\to\infty$.
We know that for any $\psi\in\text{dom}\,\hat{H}^\dagger$ both $\psi$ and $\hat{H}\psi$ have to be square integrable. Keeping this in mind we use the identity
\begin{widetext}
\begin{equation*}
	2\int_{R_0}^R d\tilde{R}\,\tilde{R}^{1-a-2b}\left(\psi^*\hat{H}\psi+ \psi\hat{H}\psi^*\right)=\left. \tilde{R}^{-a-2b}\frac{d|\psi|^2}{d\tilde{R}} \right|_{R_0}^R-2\int_{R_0}^R  d\tilde{R}\,\tilde{R}^{1-a-2b} \left( \frac{1}{\tilde{R}}\left|\frac{d\psi}{d\tilde{R}} \right|^2 - \frac{b(1+a+2b)}{\tilde{R}^3}\left|\psi \right|^2 \right),
\end{equation*}
\end{widetext}
where $0<R_0<\infty$, to argue analogously to Lemma 2.14 in \cite{GitmanSelfAdjoint} that $R^{-\frac{1}{2}}|\psi'|$ has to be square integrable near $R\to\infty$. In analogy to Lemma 2.13 in \cite{GitmanSelfAdjoint}, we can then use the identities
\begin{widetext}
\begin{align*}
	\int_{R_0}^R d\tilde{R}\,\tilde{R}^{1-a-2b}\left(\psi^*\frac{1}{\sqrt{\tilde{R}}}\frac{d\psi}{d\tilde{R}}+ \psi\frac{1}{\sqrt{\tilde{R}}}\frac{d\psi^*}{d\tilde{R}}\right)&=\left. \tilde{R}^{\frac{1}{2}-a-2b}|\psi|^2 \right|_{R_0}^R-\int_{R_0}^R  d\tilde{R}\,\tilde{R}^{1-a-2b}  \frac{\frac{1}{2}-a-2b}{\tilde{R}^\frac{3}{2}}\left|\psi \right|^2\,,\\
	2\int_{R_0}^R d\tilde{R}\,\tilde{R}^{1-a-2b}\left(\frac{1}{\sqrt{\tilde{R}}}\frac{d\psi^*}{d\tilde{R}}\hat{H}\psi+ \frac{1}{\sqrt{\tilde{R}}}\frac{d\psi}{d\tilde{R}}\hat{H}\psi^*\right)&=\left. \tilde{R}^{-\frac{1}{2}-a-2b}\left| \frac{d\psi}{d\tilde{R}}\right| ^2+b(1+a+b)\tilde{R}^{-\frac{5}{2}-a-2b}|\psi|^2 \right|_{R_0}^R\\+\int_{R_0}^R  d\tilde{R}\,&\tilde{R}^{\frac{1}{2}-a-2b} \left(  \frac{\frac{1}{2}-a-2b}{\tilde{R}^2}\left|\frac{d\psi}{d\tilde{R}} \right|^2 + \frac{b(1+a+b)(\frac{5}{2}+a+2b)}{\tilde{R}^4}\left|\psi \right|^2\right) 
\end{align*}
\end{widetext}
to deduce that for $R\to\infty$, $R^{\frac{1}{2}-a-2b}|\psi|^2\to0$
and $R^{-\frac{1}{2}-a-2b}|\psi'|^2\to0$. It directly follows that
$w(\psi,\phi)\to0$ for $R\to\infty$ and any $\psi$,
$\phi\in\text{dom}\,\hat{H}^\dagger$, meaning the $R\to\infty$ part of
\eqref{eq:w} is always fulfilled. This holds not only for
$|1+a|\geq3$, but for any factor ordering. We want to note that the
usual pathological examples for square integrable functions not
vanishing for $R\to\infty$ are excluded here by the continuity
conditions on functions in $\text{dom}\,\hat{H}^\dagger$, needed to
make the expression $\hat{H}\psi$ meaningful and the above identities
well defined due to the use of partial integration when deriving
them. 

Next we consider the boundary $R\to0$. To this end we first note that,
as mentioned previously, for a $\psi\in\text{dom}\,\hat{H}^\dagger$
with $\hat{H}^\dagger\psi=\eta$ the function $\eta$ has to be included
in $L^2(\mathbb{R}^+,R^{1-a-2b}dR)$. Using the ansatz
$\psi(R)=c_1(R)\,\phi^1_0(R)+c_2(R)\,\phi^2_0(R)$, where
$\phi^{1/2}_0$ are the zero energy stationary modes \eqref{sol_eq0},
here normalized such that $w(\phi^1_0,\phi^2_0)=1$, the above equation
can be inverted to give 
\begin{widetext}
\begin{align}
	\psi(R)&=c_1^0\,\phi^1_0(R)+c_2^0\,\phi^2_0(R)+\phi^1_0(R)\int_{R_0}^R d\tilde{R}\,\tilde{R}^{1-a-2b} \phi^2_0(\tilde{R})\eta(\tilde{R})-\phi^2_0(R)\int_{\overline{R}_0}^R d\tilde{R}\,\tilde{R}^{1-a-2b} \phi^1_0(\tilde{R})\eta(\tilde{R})\,, \label{eq:generic_phi}\\
	\psi(R)'&=c_1^0\,\phi^1_0(R)'+c_2^0\,\phi^2_0(R)'+\phi^1_0(R)'\int_{R_0}^R d\tilde{R}\,\tilde{R}^{1-a-2b} \phi^2_0(\tilde{R})\eta(\tilde{R})-\phi^2_0(R)'\int_{\overline{R}_0}^R d\tilde{R}\,\tilde{R}^{1-a-2b} \phi^1_0(\tilde{R})\eta(\tilde{R}),
\end{align}
\end{widetext}
where $c^0_1$, $c^0_2$ and $R_0$, $\overline{R}_0$ are constants, the
former complex and the latter on the real positive half line, and a
prime denotes a differentiation with respect to $R$. We can now read off how $\psi$ behaves for different factor orderings when $R\to0$.

We first note that $\phi^1_0$ is square integrable at $R=0$ for $a<2$ and at $R\to\infty$ for $a>2$, while $\phi^2_0$ is square integrable at $R=0$ for $a>-4$ and at $R\to\infty$ for $a<-4$. Hence we have to distinguish four different cases in the following, keeping in mind that we are presently only discussing the factor ordering for which $\hat{H}$ is essentially self-adjoint, $\left| 1+a\right| \geq 3$.

Let us consider $a<-4$. We choose $R_0\to\infty$ and
$\overline{R}_0=0$ such that the integrals in \eqref{eq:generic_phi} are
well defined. Using the Cauchy-Schwarz inequality we can give an
estimation for these terms in $\psi$: 
\begin{widetext}
\begin{align}
	\left| \phi^1_0(R)\int_{R}^\infty d\tilde{R}\,\tilde{R}^{1-a-2b} \phi^2_0(\tilde{R})\eta(\tilde{R})\right| &\leq \frac{2\, R^{\frac{1}{2}(4+a+2b)}}{(-1-a)\sqrt{-4-a}}\left( \int_{R}^\infty d\tilde{R}\,\tilde{R}^{1-a-2b} |\eta(\tilde{R})|^2\right)^{\frac{1}{2}} ,\label{eq:first_integral}\\
	\left|\phi^2_0(R) \int_{0}^R d\tilde{R}\,\tilde{R}^{1-a-2b} \phi^1_0(\tilde{R})\eta(\tilde{R})\right| &\leq \frac{2\, R^{\frac{1}{2}(4+a+2b)}}{(-1-a)\sqrt{2-a}}\left( \int_{0}^R d\tilde{R}\,\tilde{R}^{1-a-2b} |\eta(\tilde{R})|^2\right)^{\frac{1}{2}},\label{eq:second_integral}
\end{align}
\end{widetext}
and analogously for $\psi(R) '$, for which $\phi^{1}_0(R)$ and
$\phi^2_0(R)$ are replaced by $\phi^1_0(R)'$ and $\phi^2_0(R)'$,
decreasing the power of $R$ by one. Note that the integrals on the
right-hand side of the above estimates are bounded when $R\to0$
because $\eta$ is square integrable. 

Furthermore, we have to set $c^0_2=0$; otherwise $\psi$ is not square
integrable at $R\to0$. Plugging in the remaining terms pairwise into
$w$ and using the estimates \eqref{eq:first_integral} and
\eqref{eq:second_integral}, we can see that $w(\phi,\chi)$ always
vanishes for any functions $\phi$, $\chi$ belonging to
$\text{dom}\,\hat{H}^\dagger$ when $R\to0$ meaning that, when keeping
in mind the previous analogous result for $R\to\infty$, \eqref{eq:w}
is always fulfilled and no additional conditions are needed. 

For $a=-4$ the same conclusion holds. In this case the integral terms can respectively be estimated to behave like $R^b$ and $R^b\,\sqrt{|\ln R|}$ when approaching the boundary, which still leads to $w|_{R\to0}$ vanishing. Note that in contrast to $a<-4$ one has to choose $R_0=1$, because $\phi^2_0$ is not square integrable at either boundary.

In the case of $a>2$ we choose $R_0=0$ and $\overline{R}_0\to\infty$. Apart from minor differences concerning the signs in the prefactor and the boundaries of the integral as dictated by the aforementioned choice of $R_0$ and $\overline{R}_0$, we can estimate the integral terms as in \eqref{eq:first_integral} and \eqref{eq:second_integral}; most notably, the power of $R$ remains the same. Furthermore, we choose $c^0_1=0$. Once again none of the terms contribute to $w|_{R\to0}$. The same result emerges for the case $a=2$ ($c^0_1=0$, $R_0=0$ and $\overline{R}_0=1$), for which the integral terms can be estimated to behave like $R^{3+b}$ and $R^{3+b}\,\sqrt{|\ln R|}$.

In summary, we can say that for $|1+a|\geq3$ the domain of the essentially self-adjoint Hamiltonian is equal to $\text{dom}\,\hat{H}^\dagger$ meaning, ignoring continuity conditions, all square integrable functions $\psi$ for which $\hat{H}\psi$ is also square integrable.

Finally we have to consider $|1+a|<3$. We once again utilize \eqref{eq:first_integral} and \eqref{eq:second_integral}.
Since both $\phi^1_0$ and $\phi^2_0$ are square integrable at $R=0$ for the factor orderings in question, we can choose $R_0=\overline{R}_0=0$, and $c^0_1$, $c^0_2$ do not necessarily have to vanish. Because of $\phi^2_0$, we have to consider the case where $a=-1$ on its own. Let us first restrict ourselves to $a\neq-1$. Once again the integral terms can be estimated by \eqref{eq:first_integral} and \eqref{eq:second_integral}, with the aforementioned minor variations. The integral terms then do not contribute to $w$ as $R\to0$, but in contrast to the previously discussed factor orderings, $\phi^1_0$ and $\phi^2_0$ do.

For $a=-1$, the integral terms behave a bit differently:
\begin{widetext}
\begin{align*}
		\left| \phi^1_0(R)\int_{0}^R d\tilde{R}\,\tilde{R}^{2-2b} \phi^2_0(\tilde{R})\eta(\tilde{R})\right| &\leq \frac{2 R^{\frac{3}{2}+b}}{\sqrt{27}}\sqrt{9\ln^2R-6\ln R +2}\left( \int_{0}^R d\tilde{R}\,\tilde{R}^{2-2b} |\eta(\tilde{R})|^2\right)^{\frac{1}{2}} \,,\\
		\left| \phi^1_0(R)'\int_{0}^R d\tilde{R}\,\tilde{R}^{2-2b} \phi^2_0(\tilde{R})\eta(\tilde{R})\right| &\leq \frac{2b\, R^{\frac{1}{2}+b}}{\sqrt{27}}\sqrt{9\ln^2R-6\ln R +2}\left( \int_{0}^R d\tilde{R}\,\tilde{R}^{2-2b} |\eta(\tilde{R})|^2\right)^{\frac{1}{2}} \,,\\
		\left|\phi^2_0(R) \int_{0}^R d\tilde{R}\,\tilde{R}^{2-2b} \phi^1_0(\tilde{R})\eta(\tilde{R})\right| &\leq \frac{2}{\sqrt{3}}R^{\frac{3}{2}+b}\,|\ln R|\left( \int_{0}^R d\tilde{R}\,\tilde{R}^{2-2b} |\eta(\tilde{R})|^2\right)^{\frac{1}{2}}\,,\\
		\left|\phi^2_0(R)' \int_{0}^R d\tilde{R}\,\tilde{R}^{2-2b} \phi^1_0(\tilde{R})\eta(\tilde{R})\right| &\leq \frac{2}{\sqrt{3}}R^{\frac{1}{2}+b}\,|b\ln R+1|\left( \int_{0}^R d\tilde{R}\,\tilde{R}^{2-2b} |\eta(\tilde{R})|^2\right)^{\frac{1}{2}}.
\end{align*}
\end{widetext}
Despite the differences to previous factor orderings, the results are identical: only $\phi^1_0$ and $\phi^2_0$ contribute to $w$ as $R\to0$.

Combining the above with our previous result for the behavior of $\psi$ for $R\to\infty$, we can give an asymptotic expansion for any $\psi\in\text{dom}\,\hat{H^\dagger}$ for $R\to0$ as
\begin{equation*}
	\psi(R)=c^0_1 \phi^1_0(R) + c^0_2 \phi^2_0(R) + \tilde{\psi}_{0}(R),
\end{equation*}
where $c^0_1$, $c^0_2$ are arbitrary constants, and $\tilde{\psi}_{0}$ does not contribute to $w|_{R\to0}$. $w|_{R\to\infty}$ always vanishes, and hence there we have $\psi=\tilde{\psi}_{\infty}$. Note that the asymptotic expansion of the derivative is equal to the derivative of the asymptotic expansion above. 

This allows us to determine self-adjoint extensions for $\hat{H}$ by using Theorem 4.24 of \cite{GitmanSelfAdjoint}, where the procedure we employ below is called the `asymmetry form method'. A more pedagogical introduction to this method can be found in \cite{FulopSelfAdjoint}.

To start with, we consider the asymmetry form
\begin{equation*}
	\Delta(\psi)=W(\psi,\psi)=\left. -w(\psi,\psi)\right|_{R\to 0} ={c^0_1}^*c^0_2-{c^0_2}^*c^0_1,
\end{equation*}
where we have used that $\phi^1_0$, $\phi^2_0$ are real and normalized
such that $w(\phi^1_0,\phi^2_0)=1$. The next step is then to
diagonalize the asymmetry form, which in our case can be achieved by
defining $c_+=\frac{1}{2}\left( -c^0_1+i\,c^0_2\right)$ and
$c_-=\frac{1}{2}\left(c^0_1+i\,c^0_2\right)$ such that
\begin{align*}
	\Delta(\psi)=2i\left( |c_+|^2-|c_-|^2\right).
\end{align*}
All self-adjoint extensions of $\hat{H}$ can then be given by the condition
\begin{align}
	c_-=e^{i\theta} c_+\,,\label{eq:self-adjoint_condition}
\end{align}
where $\theta\in[0,2\pi)$.

To check whether a given $\psi\in\text{dom}\,\hat{H^\dagger}$ fulfills this condition for a given $\theta$, we need to extract the constants $c_\pm$ from the asymptotic expansion of $\psi$. To this end, we note
\begin{align*}
	\left. w(\psi,\phi^1_0)\right|_{R\to 0}&=-c^0_2\,,\\
	\left. w(\psi,\phi^2_0)\right|_{R\to 0}&=c^0_1\,.
\end{align*}
This allows us to write the condition \eqref{eq:self-adjoint_condition} as 
\begin{multline}
	\left. -(1+e^{i\theta})\,R^{2+a}~\frac{d}{d R}R^{-(1+a+b)}\psi\right|_{R\to 0}\\\left.= i(1-e^{i\theta})\,R^{-a}~\frac{d}{d R}R^{-b}\psi\right|_{R\to 0} \label{eq:boundary_condition}
\end{multline}
for $a\neq-1$, and for $a=-1$ as
\begin{multline}
\left. -(1+e^{i\theta})\,R\,\ln^2\!R~\frac{d}{d R}\frac{R^{-b}}{\ln R}\psi\right|_{R\to 0}\\\left.= i(1-e^{i\theta})\,R~\frac{d}{d R}R^{-b}\psi\right|_{R\to 0}\,.
\end{multline}
Finally we note that $\theta$ can, of course, be chosen differently for each factor ordering. We thus change $\theta$ according to $\theta\to-\theta+\pi$ for $a>-1$, allowing us to rewrite \eqref{eq:boundary_condition} as
\begin{multline}
  \label{b-condition}
\left. -(1+e^{i\theta})\,R^{1-|1+a|}~\frac{d}{d R}R^{-\frac{1}{2}(1+a-|1+a|+2b)}\psi\right|_{R\to 0}\\\left.= i(1-e^{i\theta})\,R^{1+|1+a|}~\frac{d}{d R}R^{-\frac{1}{2}(1+a+|1+a|+2b)}\psi\right|_{R\to 0} .
\end{multline}
It turns out that this form of the boundary conditions works best for
our stationary modes. This concludes our discussion of the
self-adjoint extensions of the Hamiltonian. 

\section{}\label{sec:boundary_conditions_full}
We want to enforce the boundary conditions
\eqref{eq:condition_a_not_-1} and \eqref{eq:condition_a_-1}, which correspond to the different self-adjoint extensions of the Hamiltonian for the positive energy stationary modes $\phi^1_E$ and $\phi^2_E$. Recall that only factor orderings with $|1+a|<3$ are relevant here. First we will consider $a\neq-1$.

We start with $\phi^1_E$ and compute
\begin{widetext}
	\begin{align*}
	R^{1-|1+a|}&\frac{d}{d
		R}R^{-\frac{1}{2}(1+a-|1+a|+2b)}\phi^1_{E}\nonumber\\&=-\sqrt{2E}R^{\frac{1}{2}\left( 
		3-|1+a|\right) }\left(\cos\!\left(\tfrac{\pi}{3}|1+a|\right)
	J_{1-\frac{1}{3}\left| 1+a\right|
	}\!\left(\tfrac{2}{3}\sqrt{2E}R^{\frac{3}{2}}\right)+
	\sin\!\left(\tfrac{\pi}{3}|1+a|\right) Y_{1-\frac{1}{3}\left|
		1+a\right| }\!\left(\tfrac{2}{3}\sqrt{2E}R^{\frac{3}{2}}
	\right)\right) \\& 
	\overset{R\to0}{\sim}-\frac{3\cos\!\left(\tfrac{\pi}{3}|1+a|\right) }{\Gamma\!\left(2-\tfrac{1}{3}|1+a| \right)}\left(\tfrac{1}{3}\sqrt{2E} \right)^{2-\tfrac{1}{3}\left|1+a \right|} R^{3-|1+a|}+\tfrac{3}{\pi}\sin\!\left(\tfrac{\pi}{3}|1+a|\right)\Gamma\!\left(1-\tfrac{1}{3}|1+a| \right)\left(\tfrac{1}{3}\sqrt{2E} \right)^{\tfrac{1}{3}\left|1+a \right|},\\
	R^{1+|1+a|}&\frac{d}{d
		R}R^{-\frac{1}{2}(1+a+|1+a|+2b)}\phi^1_{E}=-\sqrt{2E}R^{\frac{1}{2}\left(
		3+|1+a|\right) }~J_{1+\frac{1}{3}\left| 1+a\right|
	}\!\left(\tfrac{2}{3}\sqrt{2E}R^{\frac{3}{2}} \right) 
	\overset{R\to0}{\sim}-\frac{3 \left(\tfrac{1}{3}\sqrt{2E}
		\right)^{2+\tfrac{1}{3}\left|1+a
			\right|}}{\Gamma\!\left(2+\tfrac{1}{3}|1+a| \right)}
	R^{3+|1+a|},
	\end{align*}
\end{widetext}
where we have used several well known identities of the Bessel functions and their derivatives, which can be found e.g. in \cite{DLMF}, along with their asymptotic behavior. Inserting $\phi^1_E$ on its own into \eqref{eq:condition_a_not_-1} would thus lead to $e^{i\theta}+1=0$. $\phi^1_E$ is hence viable for $\theta=\pi$, but for other self-adjoint extensions we have to consider more general linear combinations of the two modes.

For $\phi^2_E$ we proceed along the same lines as for $\phi^1_E$ and compute
\begin{widetext}
	\begin{align*}
	R^{1-|1+a|}&\frac{d}{d R}R^{-\frac{1}{2}(1+a-|1+a|+2b)}\phi^2_{E}\nonumber\\&=\sqrt{2E}R^{\frac{1}{2}\left( 3-|1+a|\right) }\left(\sin\!\left(\tfrac{\pi}{3}|1+a|\right) J_{1-\frac{1}{3}\left| 1+a\right| }\!\left(\tfrac{2}{3}\sqrt{2E}R^{\frac{3}{2}}\right)- \cos\!\left(\tfrac{\pi}{3}|1+a|\right) Y_{1-\frac{1}{3}\left| 1+a\right| }\!\left(\tfrac{2}{3}\sqrt{2E}R^{\frac{3}{2}} \right)\right) \\&
	\overset{R\to0}{\sim}~\frac{3\sin\!\left(\tfrac{\pi}{3}|1+a|\right) }{\Gamma\!\left(2-\tfrac{1}{3}|1+a| \right)}\left(\tfrac{1}{3}\sqrt{2E} \right)^{2-\tfrac{1}{3}\left|1+a \right|} R^{3-|1+a|}+\tfrac{3}{\pi}\cos\!\left(\tfrac{\pi}{3}|1+a|\right)\Gamma\!\left(1-\tfrac{1}{3}|1+a| \right)\left(\tfrac{1}{3}\sqrt{2E} \right)^{\tfrac{1}{3}\left|1+a \right|},\\
	R^{1+|1+a|}&\frac{d}{d R}R^{-\frac{1}{2}(1+a+|1+a|+2b)}\phi^2_{E}=-\sqrt{2E}R^{\frac{1}{2}\left( 3+|1+a|\right) }~Y_{1+\frac{1}{3}\left| 1+a\right| }\!\left(\tfrac{2}{3}\sqrt{2E}R^{\frac{3}{2}} \right)\\&
	\overset{R\to0}{\sim}\frac{3}{\pi} \Gamma\!\left(1+\tfrac{1}{3}|1+a| \right) \left(\tfrac{1}{3}\sqrt{2E} \right)^{-\tfrac{1}{3}\left|1+a \right|},
	\end{align*}
\end{widetext}

On its own, $\phi^2_E$ would only be able to fulfill \eqref{eq:condition_a_not_-1} for a single specific energy, but since it is not square integrable, it does not admit an interpretation as a bound state. As noted above, only a specific linear combination $A\phi^1_E+B\phi^2_E$, $A\neq0$, is permissible under \eqref{eq:condition_a_not_-1}:
\begin{multline*}
-(1+e^{i\theta})\,\Gamma\!\left(1-\tfrac{1}{3}|1+a| \right)\left(\tfrac{1}{3}\sqrt{2E} \right)^{\tfrac{1}{3}\left|1+a \right|}\\\times\left(A\, \sin\!\left(\tfrac{\pi}{3}|1+a|\right) + B\, \cos\!\left(\tfrac{\pi}{3}|1+a|\right)\right) \\=
i(1-e^{i\theta})\,\Gamma\!\left(1+\tfrac{1}{3}|1+a| \right) \left(\tfrac{1}{3}\sqrt{2E} \right)^{-\tfrac{1}{3}\left|1+a \right|}B.
\end{multline*}
For $\theta=\pi$ the above implies $B=0$, and hence we see that $\phi^1_E$ and only $\phi^1_E$ is viable for this self-adjoint extension. With $\theta\neq\pi$ we continue and arrive at

\begin{multline*}
A\, \sin\!\left(\tfrac{\pi}{3}|1+a|\right) + B\, \cos\!\left(\tfrac{\pi}{3}|1+a|\right) \\=
-\tan\tfrac{\theta}{2}\,\frac{\Gamma\!\left(1+\tfrac{1}{3}|1+a| \right)}{\Gamma\!\left(1-\tfrac{1}{3}|1+a| \right)} \left(\tfrac{1}{3}\sqrt{2E} \right)^{-\tfrac{2}{3}\left|1+a \right|}B,
\end{multline*}
and hence the positive energy stationary mode permitted by \eqref{eq:condition_a_not_-1} is

\begin{multline*}
	-\tan\tfrac{\theta}{2}\,\frac{\Gamma\!\left(1+\tfrac{1}{3}|1+a| \right)}{\Gamma\!\left(1-\tfrac{1}{3}|1+a| \right)} \left(\tfrac{1}{3}\sqrt{2E} \right)^{-\tfrac{2}{3}\left|1+a \right|}\,\phi^1_E\\ -\cos\!\left(\tfrac{\pi}{3}|1+a|\right)\,\phi^1_E  +\sin\!\left(\tfrac{\pi}{3}|1+a|\right)\,\phi^2_E.
\end{multline*}

Finally we want to consider the case $a=-1$. Plugging $A\phi^1_E+B\phi^2_E$ into \eqref{eq:condition_a_-1} and a straightforward calculation leads to
\begin{equation*}
	(1+e^{i\theta}) \left(A+\frac{2B}{\pi}\ln\!\left( \tfrac{2}{3}\sqrt{2E}\right) \right) \\= i(1-e^{i\theta})\frac{3B}{\pi}.
\end{equation*}
As is apparent, for $\theta=\pi$ we once again have as our permitted mode $\phi^1_E$, and for other factor orderings
\begin{equation*}
	\left(\tfrac{3}{\pi}\,\tan\tfrac{\theta}{2}-\tfrac{2}{\pi}\,\ln\left( \tfrac{2}{3}\sqrt{2E}\right) \right) \phi^1_E+\phi^2_E.
\end{equation*}

\section{}\label{sec:app_R0_full}
The expectation value of the minimal radius for the full wave packet \eqref{eq:general_wavepacket} can be computed as
\begin{widetext}
	\begin{multline*}
	\overline{R}(\tau=0) = \int_0^\infty dR~R^{1-a-2b}~R~\left| \Psi(R,\tau=0)\right| ^2=\lambda^\frac{1}{3}\frac{ 2^{\kappa +\frac{1}{3}}\pi}{3^\frac{1}{3}}\frac{\csc \!\left(\frac{\pi}{6} (\left| a+1\right| -3 \kappa +2)\right) \Gamma\! \left(\frac{\left| a+1\right|}{6} +\frac{\kappa }{2}+1\right) }{ \Gamma\! \left(\frac{2}{3}\right) \Gamma (\kappa +1) \Gamma\! \left(\frac{\left| a+1\right|}{6} -\frac{\kappa }{2}\right)}\\\times\left[\Gamma\! \left(\tfrac{\left| a+1\right|}{3} +\tfrac{4}{3}\right)\Gamma\! \left(\tfrac{\left| a+1\right|}{6} -\tfrac{ \kappa}{2} \right)~ _3\tilde{F}_2\!\left(\tfrac{4}{3},\tfrac{\left| a+1\right|}{6} +\tfrac{\kappa}{2} +1,\tfrac{\left| a+1\right|}{3} +\tfrac{4}{3};\tfrac{\left| a+1\right|}{6} -\tfrac{\kappa}{2} +\tfrac{4}{3},\tfrac{\left| a+1\right| }{3} +1;-1\right) \right. \\\left. +3 ~\Gamma\! \left(\tfrac{2}{3}\right)\Gamma\! \left(\kappa +\tfrac{2}{3}\right)~_3\tilde{F}_2\!\left(\kappa +\tfrac{2}{3},-\tfrac{\left| a+1\right|}{6}+\tfrac{\kappa}{2} +1,\tfrac{\left| a+1\right|}{6}+\tfrac{\kappa}{2} +1;-\tfrac{\left| a+1\right|}{6}+\tfrac{\kappa}{2}  +\tfrac{2}{3},\tfrac{\left| a+1\right|}{6}+\tfrac{\kappa}{2}  +\tfrac{2}{3};-1\right)\right],
	\end{multline*}
\end{widetext}
where $_3\tilde{F}_2$ are regularized hypergeometric functions. The
function $g(a,\kappa)$ follows from comparison of the above with the
expression $\overline{R}(\tau=0)=\lambda^\frac{1}{3}\,g(a,\kappa)$. It
is obvious that it intricately depends on both $a$ and $\kappa$.

\section{}\label{sec:app_lifetime}
The full expression for the probability for the transition from
collapse to grey hole state, as well as from collapse to expansion is
rather complicated,
\begin{widetext}
	\begin{multline}
	\frac{P_{{\rm C}\to {\rm E}}}{P_{{\rm C}\to
            {\rm GH}}}=\frac{\int_{-\infty}^{-\tau_{\rm
              AH}}d\tau_-\int_{\tau_{\rm
              AH}}^{\infty}d\tau_+~W(\tau_-,\tau_+)}{\int_{-\infty}^{-\tau_{\rm
              AH}}d\tau_-\int_{-\tau_{\rm AH}}^{\tau_{\rm
              AH}}d\tau_+~W(\tau_-,\tau_+)}\\=\frac{ \frac{\frac{2}{3}
            \left| 1+a\right|}{\frac{2}{3} \left| 1+a\right| +1}  \,
          _2F_1\!\left(\frac{1}{2},1;\frac{\left| 1+a\right|
            }{3}+\frac{3}{2};-\frac{\lambda ^2}{4 \tau _{\rm
                AH}^2}\right)-1}{1-\left(\frac{4 \tau _{\rm
                AH}^2}{\lambda ^2}+1\right)^{\frac{\left| 1+a\right|
            }{3}} \left(1+\frac{2 \sqrt{\pi } \tau _{\rm AH} \Gamma
              \left(\frac{\left|1+a\right|
                }{3}+\frac{1}{2}\right)}{\lambda \Gamma
              \left(\frac{\left| 1+a\right| }{3}\right) }-\frac{8 \tau
              _{\rm AH}^2 \left|1+a\right|  }{3 \lambda ^2} \,
            _2F_1\!\left(\frac{1}{2},\frac{\left|1+a\right|
              }{3}+1;\frac{3}{2};-\frac{4 \tau _{\rm AH}^2}{\lambda
                ^2}\right)\right)}\,.\label{eq:lifetime_prob_1} 
	\end{multline}
\end{widetext}

Keeping in mind that $\frac{\tau_{\rm AH}}{\lambda}$ is roughly proportional to $\overline{E}^2$ for $a$ fixed, we can approximate the result for high energies. To this end, we note that asymptotically the Gauss hypergeometric function $_2F_1$ behaves like \cite{DLMF}
\begin{align*}
_2F_1\left(a,b;c;z\right)&\approx\frac{\Gamma (c) \Gamma (b-a)}{\Gamma (b) \Gamma (c-a)}(-z)^{-a}\nonumber\\&\quad+\frac{\Gamma (c) \Gamma (a-b)}{\Gamma (a) \Gamma (c-b)}(-z)^{-b}&\text{for }|z|\to\infty,
\end{align*}
\begin{align*}
_2F_1\left(a,b;c;z\right)&\approx 1+\frac{ab}{c}z&\text{for }|z|\to 0 .
\end{align*}
Applying this to \eqref{eq:lifetime_prob_1} above gives
\begin{equation*}
	\frac{P_{{\rm C}\to {\rm E}}}{P_{{\rm C}\to {\rm GH}}}\approx\frac{\left(2\frac{\tau_{\rm AH}}{\lambda}\right)^{-\frac{2}{3}\left|1+a \right|} }{\frac{2}{3}\left|1+a \right|+1}.
\end{equation*}

The same approximation can be applied to the transition probability from grey hole to expanding state,
\begin{widetext}
	\begin{align}
	P_{{\rm GH}\to {\rm E}}&=\frac{\int_{-\tau_{\rm AH}}^{\tau_{\rm AH}}d\tau_-\int_{\tau_{\rm AH}}^{\infty}d\tau_+~W(\tau_-,\tau_+)}{\int_{-\tau_{\rm AH}}^{\tau_{\rm AH}}d\tau_-\int_{-\infty}^{\infty}d\tau_+~W(\tau_-,\tau_+)}=\frac{1}{2}+\frac{ \Gamma \!\left(\tfrac{1}{3}\left| a+1\right| \right)}{4 \sqrt{\pi } \Gamma \!\left(\tfrac{1}{3}\left| a+1\right| +\frac{1}{2}\right)}\frac{\lambda}{\tau_{\rm AH}}\left(1-\left(1+\frac{4 \tau _{\rm AH}^2}{\lambda^2}\right)^{-\tfrac{1}{3}\left| a+1\right| }\right)-\nonumber\\&\quad-\frac{2 \Gamma \!\left(\tfrac{1}{3}\left| a+1\right| +1\right)}{\sqrt{\pi } \Gamma\! \left(\tfrac{1}{3}\left| a+1\right| +\frac{1}{2}\right)}\frac{\tau_{\rm AH}}{\lambda} \, _2F_1\!\left(\tfrac{1}{2},\tfrac{1}{3}\left| a+1\right| +1;\tfrac{3}{2};-\tfrac{4 \tau _{\rm AH}^2}{\lambda ^2}\right)\\
	&\approx\frac{ \Gamma\! \left(\tfrac{1}{3}\left| a+1\right| \right)}{4 \sqrt{\pi } \Gamma\! \left(\tfrac{1}{3}\left| a+1\right| +\frac{1}{2}\right)}\frac{\lambda}{\tau_{\rm AH}}\,.\label{eq:lifetime_prob_2}
	\end{align}
\end{widetext}
Note that this approximation for high energies only applies when $a\neq-1$, otherwise \eqref{eq:lifetime_prob_2} behaves like $\frac{\lambda}{\tau_{\rm AH}}\,\ln\!\left(2\frac{\tau_{\rm AH}}{\lambda}\right) $.



\end{document}